\documentclass[twocolumn,aps,prd,preprintnumbers,nofootinbib]{revtex4-2}
\usepackage{graphicx}
\usepackage{amsmath}
\usepackage{amssymb}
\usepackage{amsfonts}
\usepackage{hyperref}
\usepackage{url}
\usepackage{color}
\usepackage{bm}

\newcommand{\be}{\begin{equation}}
\newcommand{\ee}{\end{equation}}
\newcommand{\ba}{\begin{eqnarray}}
\newcommand{\ea}{\end{eqnarray}}
\newcommand{\nn}{\nonumber}

\renewcommand{\[}{\begin{equation}}
\renewcommand{\]}{\end{equation}}

\def\lcdm{$\Lambda$CDM }

\usepackage{array}
\makeatother

\begin{document}

\preprint{IFT-UAM/CSIC-20-189}

\title{Machine Learning and cosmographic reconstructions of quintessence and the Swampland conjectures}

\author{Rub\'{e}n Arjona}
\email{ruben.arjona@uam.es}

\author{Savvas Nesseris}
\email{savvas.nesseris@csic.es}

\affiliation{Instituto de F\'isica Te\'orica UAM-CSIC, Universidad Auton\'oma de Madrid,
Cantoblanco, 28049 Madrid, Spain}

\date{\today}

\begin{abstract}
We present model independent reconstructions of quintessence and the Swampland conjectures (SC) using both Machine Learning (ML) and cosmography. In particular, we demonstrate how the synergies between theoretical analyses and ML can provide key insights on the nature of dark energy and modified gravity. Using the Hubble parameter $H(z)$ data from the cosmic chronometers we find that the ML and cosmography reconstructions of the SC are compatible with observations at low redshifts. Finally, including the growth rate data $f\sigma_8(z)$ we perform a model independent test of modified gravity cosmologies through two phase diagrams, namely $H-f\sigma_8$ and $\eta-f\sigma_8$, where the anisotropic stress parameter $\eta$ is obtained via the $E_g$ statistics, which is related to gravitational lensing data. While the first diagram is consistent within the errors with the $\Lambda$CDM model, the second one has a $\sim 2\sigma$ deviation of the anisotropic stress from unity at $z\sim 0.3$ and a $\sim 4\sigma$ deviation at $z\sim 0.9$, thus pointing toward  mild deviations from General Relativity, which could be further tested with upcoming large-scale structure surveys.
\end{abstract}
\maketitle


\section{Introduction}
The observed accelerated expansion of the Universe at late times is one of the most challenging enigmas in theoretical physics. Observations over the last two decades or so, using Type Ia Supernovae, the cosmic microwave background (CMB) anisotropies and large scale structure (LSS) data imply the existence of a repulsive force that dominates over gravity on cosmological (large) scales. In the context of General Relativity (GR), this suggests the presence of an energy-momentum component with a negative equation of state, referred to as Dark Energy (DE). So far, with the current available data, the best phenomenological explanation for this accelerated expansion is the cosmological constant $\Lambda$ \cite{Aghanim:2018eyx}, which behaves as a uniform vacuum energy spread over all space.

However, the cosmological constant introduces extreme fine-tuning since its value is much smaller than what is predicted by quantum field theories \cite{Weinberg:1988cp,Carroll:2000fy} and also suffers from the coincidence problem. This has given rise to a plethora of different DE models based on ad-hoc ideal fluids or new scalar fields that mediate the force between particles ranging from canonical scalar fields \cite{Ratra:1987rm,Wetterich:1987fm,Caldwell:1997ii}, scalar fields with a generalized kinetic terms \cite{ArmendarizPicon:2000dh,ArmendarizPicon:2000ah}, non-minimal couplings \cite{Uzan:1999ch,Perrotta:1999am,Riazuelo:2001mg} or coupled DE models \cite{Dent:2008vd} in addition to GR.

One of the more well known candidates for DE is quintessence, which is described by a slowly rolling scalar field leading to an accelerated expansion \cite{Amendola:2015ksp}. Thus, the scalar field can control the fate of the early and late Universe by dominating its energy density and posing as a source of DE respectively, see for example Ref.~\cite{Tsujikawa:2013fta} for a review.

Another viable alternative to DE models can be found through covariant modifications of GR, known as Modified Gravity (MG) theories. These are motivated by high energy physics (Quantum Gravity and String Theory) and have specific signatures, e.g. a time and scale dependent Newton's constant, a different evolution of the matter density perturbations or specific patterns in the emission of gravitational waves, compared to GR. From this point of view, GR is seen as an effective low-energy theory that, as we increase the energy scale, higher order corrections are needed \cite{tHooft:1974toh}.

In view of the plethora of DE and MG models, there has been an effort to provide a unified framework which encloses several of them like the Effective Field Theory (EFT) approach \cite{Gubitosi:2012hu,Hu:2013twa} or the Effective Fluid approach (EFA) \cite{Arjona:2018jhh,Arjona:2019rfn,Arjona:2020gtm}. Despite this large variety of models, current Bayesian analyses of astrophysical measurements indicate that the standard cosmological model which contains the cosmological constant $\Lambda$ and a cold dark matter component (CDM) \cite{Peebles:2002gy} outperforms all other models \cite{Heavens:2017hkr}.

For about a century theoretical physicists have been on the quest to develop a theory of quantum gravity which could encompass the assumptions of Einstein's theory of GR with those of quantum field theory. Although GR has demonstrated to have a very high predictive power below the Planck scale, its quantization its troublesome since it is renormalizable only at one loop \cite{Birrell:1982ix}, thus, it is believed that GR could be the low energy limit of the more fundamental higher energy theory.

There is an ongoing search to distinguish effective quantum field theories that can potentially arise within UV-complete quantum gravity theories (the Landscape) from those that cannot (the Swampland). In this regard, although not rigorously proven in string theory, some conjectures have been considered to discern the Swampland from the landscape. The two proposed Swampland criteria that we will consider (which we will define as SC1 and SC2) refer to the constraints on the field range of a scalar field $\phi$ defined by an effective field theory and to the slope of the potential of such fields respectively. In reduced Planck units these conjectures are defined as
\begin{enumerate}
  \item SC1: The scalar field net excursion has to satisfy $\frac{|\Delta \phi|}{M_\textrm{pl}}<\Delta \sim O(1)$ \cite{Ooguri:2006in}.
  \item SC2: There is a lower bound for the gradient of the scalar field potential $M_\textrm{pl}\left|\nabla_{\phi} V\right| / V>c \sim \mathcal{O}(1)$ in any consistent theory of gravity when $V>0$ \cite{Obied:2018sgi},
\end{enumerate}
where $\Delta$ and $c$ are positive constants of order one and the reduced Planck mass is $M_\textrm{pl}=1/\sqrt{8\pi G}$. The second Swampland criterion is violated in the $\Lambda$\text{CDM} model, since a positive cosmological constant or being at the minimum of a potential with positive energy density violates the bound \cite{Agrawal:2018own}, thus a rolling scalar field potential, i.e a quintessence model would be required. Hence, if the data supports the second Swampland criterion, it would imply hints for deviations of the $\Lambda$\text{CDM} model.

The Swampland criteria aim to find constructions that are compatible with a quantum theory of gravity and it has been found that specific quintessence models can satisfy the Swampland criteria at late times \cite{Agrawal:2018own}. In Ref.~\cite{Elizalde:2018dvw} the authors used Gaussian Processes to reconstruct the form of the potential from the $H(z)$ data, finding hints of invalidating the Swampland criteria, while a similar analysis was performed in \cite{Yang:2020jze}. In Refs.~\cite{Colgain:2019joh,Banerjee:2020xcn} it was found that quintessence models and current data prefer a lower value of $H_0$ than the $\Lambda$\text{CDM} model, thus providing robust test of the Swampland conjectures. Other analyses on the other hand, have found that string-inspired quintessence models with exponential potentials are ruled out by observations and that Swampland conjectures are in tension with viable single-field quintessence models \cite{Akrami:2018ylq,Raveri:2018ddi}. It was also proposed though that this issue might be resolved with multi-field models \cite{Garg:2018reu,Akrami:2020zfz}. See also Ref.~\cite{Heisenberg:2018yae} for the implications of the swampland conjectures on dark energy.

On the other hand, our motivation for using cosmography and Machine Learning (ML), both being model independent techniques, is because choosing a specific model can lead to model bias, which in turn would affect the conclusions drawn about fundamental physics. ML algorithms can help to remove biases due to choosing a priori a specific defined model and they are also ideal for events that are not well understood such as dark energy, dark matter or modifications of gravity. Another advantage is that we reconstruct the data without making assumptions on flatness or a dark energy model.

Here we use a particular ML method known as the genetic algorithms (GA), which can be defined as a stochastic search approach. However, in our analysis we will use both cosmography and the GA so as to compare the two methods and examine which one provides better constraints given the current data. In particular, we will focus on quintessence as an example of our approach and using the latest compilation of the Hubble parameter $H(z)$ and the growth rate data $f\sigma_8(z)$ we analyze the cosmological implications on two Swampland criteria providing constraints both via Machine Learning and cosmography. For the former approach, we reconstruct the Hubble function $H(z)$ and $f\sigma_8(z)$ using the GA, while with the later method we can express the Swampland conjectures solely via the cosmographic parameters.

Finally, in order to test and search for deviations from GR we use our ML reconstructions to analyze two phase diagrams, $H-f\sigma_8$ and $\eta-f\sigma_8$, where $\eta$ is the anisotropic stress parameter. This conjoined diagrams have the asset of helping to break degeneracies between observations that are geometrical against those that come from gravitational effects and makes clearer even visually which redshift ranges should be the target of future surveys to discriminate among the plethora of DE and MG models. This approach has been used for different comparison of models, see for example Refs.~\cite{Sagredo:2018rvc,Moresco:2017hwt,Basilakos:2017rgc,Linder:2016xer,Matsumoto:2020kdu}.

Our paper is organized as follows: In Sec.~\ref{sec:theory} we present the theoretical framework including the quintessence reconstruction and the cosmographic expansion. In Sec.~\ref{sec:data} we describe the data used in our analysis and in Sec.~\ref{sec:GA} we outline our ML method, the Genetic Algorithms (GA). Then in Sec.~\ref{sec:results} we set out our results and in Sec.~\ref{sec:diagrams} we provide two phase diagrams derived through our ML reconstructions. Finally in Sec.~\ref{sec:conclusions} we present our conclusions.
\section{Theory \label{sec:theory}}
Here we present some theoretical aspects of our analysis related to the reconstruction of quintessence and the cosmographic expansion.

\subsection{Quintessence reconstruction}
At late times, the Friedmann equations including  quintessence can be written as
\ba
H^{2} &=&\frac{8\pi G}{3} \left(\rho_{m}+\frac{1}{2} \dot{\phi}^{2}+V(\phi)\right), \\
\dot{H} &=&-4 \pi G\left(\rho_{m}+\dot{\phi}^{2}\right),
\ea
where $H\equiv\frac{\dot{a}}{a}$, for $a=\frac1{1+z}$ and after setting $x\equiv1+z$ they can be solve for the potential and the kinetic terms and be rewritten as \cite{Sahni:2006pa}
\ba
\frac{8 \pi G}{3 H_{0}^{2}} V(x) &=&\frac{H(x)^{2}}{H_{0}^{2}}-\frac{x}{6 H_{0}^{2}} \frac{d (H(x)^{2})}{d x}-\frac{1}{2} \Omega_{\mathrm{m},0} x^{3}\label{eq:Vx0},~~ \\
\frac{8 \pi G}{3 H_{0}^{2}}\left(\frac{d \phi}{d x}\right)^{2} &=&\frac{2}{3 H_{0}^{2} x} \frac{d \ln H}{d x}-\frac{\Omega_{\mathrm{m},0} x}{H^{2}}\label{eq:phix0}.
\ea
It is more convenient to rescale all variables and use dimensionless quantities, which can be done for example by introducing the Planck mass $M_\textrm{pl}\equiv\sqrt{\frac{\hbar c}{8 \pi G}}=\sqrt{\frac{1}{8 \pi G}}$ in natural units ($\hbar=c=1$) and the fact that the critical density is $\rho_c=\frac{3H_0^2}{8\pi G}$. Then we can make the redefinitions
\ba
E(z)&\equiv&H(z)/H_0, \nn \\
\tilde{\phi}(z)&\equiv&\frac{\phi(z)}{\sqrt{3}M_\textrm{pl}},\nn \\
\tilde{V}(z)&\equiv&\frac{V(z)}{\rho_c},
\ea
and rewrite the reconstruction equations for the scalar field as
\ba
\tilde{V}(x) &=&E(x)^2-\frac{x}{6} \frac{d (E(x)^{2})}{d x}-\frac{1}{2} \Omega_{\mathrm{m},0} x^{3}\label{eq:Vx}, \\
\left(\frac{d \tilde{\phi}}{d x}\right)^{2} &=&\frac{2}{3x} \frac{d \ln E}{d x}-\frac{\Omega_{\mathrm{m},0} x}{E(x)^{2}}\label{eq:phix}.
\ea

To reconstruct the potential we then integrate Eq.~(\ref{eq:phix}) to determine $\tilde{\phi}(x)$ up to a constant, then we write $x$ as a function of $\tilde{\phi}$ i.e $x(\tilde{\phi})$ and insert it in  Eq.~(\ref{eq:Vx}) to find the potential in terms in the scalar field $\tilde{V}(\tilde{\phi})$. For the ML approach the function $H(x)$ and the parameters $H_0$ and $\Omega_{\mathrm{m},0}$ will be given by the GA fits to the data, as described in Section \ref{sec:data}, while in the case of cosmography we will determine the function $H(x)$ from the cosmographic reconstruction and we will assume a Planck 2018 prior on $\Omega_{\mathrm{m},0}$.

Note that using the aforementioned equations one may try to reconstruct any DE model, e.g. the constant equation of state $w$=const model or other parameterized $w(z)$ models \cite{Scherrer:2015tra}.

\subsection{Cosmography \label{sec:cosmo}}
Cosmography is a model independent series expansion in terms of the redshift $z$ that relates the cosmological quantities, such as the Hubble parameter and luminosity distance, to a set of cosmographic coefficients defined as the nth derivative of the scale factor  \cite{Visser:2004bf,Capozziello:2008qc,Capozziello:2019cav,Aviles:2012ay}:
\ba
H & \equiv & \frac{1}{a} \frac{d a}{d t}, \hspace{12mm} q \equiv -\frac{1}{a H^{2}} \frac{d^{2} a}{d t^{2}}, \\
j & \equiv & \frac{1}{a H^{3}} \frac{d^{3} a}{d t^{3}}, \hspace{5.5mm} s  \equiv \frac{1}{a H^{4}} \frac{d^{4} a}{d t^{4}}, \\
l & \equiv & \frac{1}{a H^{5}} \frac{d^{5} a}{d t^{5}}, \hspace{4mm} m  \equiv \frac{1}{a H^{6}} \frac{d^{6} a}{d t^{6}}.
\ea
With simple algebra we can relate these quantities, evaluated today, i.e. at $z=0$, to the series expansions of the Hubble parameter and the luminosity distance. For example, following Ref.~\cite{Visser:2004bf} we find that the luminosity distance in a flat Universe $(\Omega_k=0)$ can be written up to fifth order in redshift as:
\ba
d_L(z)&=& \frac{c}{H_0}\Big[z + \frac12 (1-q_0) z^2 + \frac16 (-1 - j_0 + q_0 + 3 q_0^2) z^3 \nn \\
&+& \frac{1}{24} (2 + 5 j_0 (1 + 2 q_0) - q_0 (2 + 15 q_0 (1 + q_0)) + s_0) z^4 \nn \\
&+& \frac{1}{120}\big((-6+10 j_0^2-l_0-j_0(27+5q_0(22+21q_0))\nn\\
&+&3 q_0(2+q_0(27+5q_0(11+7q_0))-5s_0)-11s_0)\big)z^5\nn \\
&+& \mathcal{O}(z^6)\Big],\label{eq:cosmodL}
\ea
while by inverting the equation of the luminosity distance that relates it to the Hubble parameter for a flat Universe, i.e. $d_L(z)=\frac{c}{H_0}(1+z)\int_0^z\frac{1}{H(u)/H_0}du$, and solving for $H(z)$ we find
\ba
H(z)/H_0&=& 1 + (1 + q_0) z + \frac12 (j_0 - q_0^2) z^2 \nn \\
&+&\frac16 \left(3 q_0^2 (1 + q_0) - j_0 (3 + 4 q_0) - s_0\right) z^3 \nn \\
&+& \frac{1}{24} \big(-4j_0^2+l_0-3 q_0^2 (4+q_0 (8+5 q_0))\nn \\
&+&j_0 (12+q_0 (32+25 q_0))+(8+7 q_0) s_0\big) z^4\nn \\
&+&\mathcal{O}(z^5).\label{eq:cosmoH}
\ea

Note that going from the luminosity distance given by Eq.~\eqref{eq:cosmodL} to the Hubble parameter given by Eq.~\eqref{eq:cosmoH}, implies the use of differentiation and the presence of a term $1+z$, both of which reduce the order of the polynomial from fifth order to only fourth. This reduction of the polynomial will also be observed later on, when we derive the potential as a function of the scalar field and the cosmographic parameters.

Furthermore, it should be noted that there is an issue related to the convergence of the truncation order of the cosmographic series and the redshift range of the data. In Refs.~\cite{Cattoen:2007id,Lazkoz:2013ija,Guimaraes:2010mw} it has been suggested that the variable $y=\frac{z}{1+z}$ avoids the aforementioned convergence issues and is more suitable for parameterizing cosmological distances, where now $y$ lays in the redshift interval $\left[0,1\right]$ which encloses the range of all possible observations. 

Using Eq.~\eqref{eq:cosmoH} we can now use the quintessence reconstruction set of equations given by Eqs.~\eqref{eq:Vx}-\eqref{eq:phix} to relate the cosmographic parameters to the potential $V(\phi)$, which after some simple algebra can be written in terms of the redshift $z$ as
\ba
\tilde{V}(z)&=&\frac{1}{6}\left(4-2q_0-3\Omega_{\mathrm{m}, 0}\right)+\frac{1}{6}\left(8-2j_0+6q_0-9\Omega_{\mathrm{m},0}\right)z\nn\\
&+&\frac{1}{6}\left(4+8q_0+j_0(4+q_0)+s_0-9\Omega_{\mathrm{m},0}\right)z^2\nn\\
&+&\frac{1}{18}(j_0^2-l_0-9\Omega_{\mathrm{m},0}-j_0q_0(7+3q_0)-7s_0-3q_0s_0)z^3\nn \\
&+&\mathcal{O}(z^4),
\ea
and the derivative of the scalar field $\left(\frac{d\tilde{\phi}'}{dz}\right)^2$ as
\ba
\tilde{\phi}'(z)^2&=&\frac{2(1+q_0)}{3}-\Omega_{\mathrm{m},0}+\frac{1}{3}\big(-4+2j_0-4q_0^2 \nn\\
&+& 6q_0(\Omega_{\mathrm{m},0}-1)+3\Omega_{\mathrm{m},0}\big)z
+\frac{1}{3}\big(6-s_0-3\Omega_{\mathrm{m},0} \nn\\
&+& j_0(-8-7q_0+3\Omega_{\mathrm{m},0})+4q_0(3+4q_0+2q_0^2 \nn\\
&-& 3(1+q_0)\Omega_{\mathrm{m},0})\big)z^2+\mathcal{O}(z^3).
\ea
Solving for $\tilde{\phi}$ by integrating the kinetic term over the redshift, will give two branches as
\be
\tilde{\phi}(z)=\tilde{\phi}_0+\epsilon \int_0^z\sqrt{\tilde{\phi}'(u)^2}du,
\ee
where $\epsilon=\pm1$. We can then express the potential in terms of the cosmographic parameters and the scalar field $\tilde{\phi}$ as
\ba
\tilde{V}(\tilde{\phi})&=&\tilde{V}_0+\tilde{V}_1(\tilde{\phi}-\tilde{\phi}_0)+\tilde{V}_2(\tilde{\phi}-\tilde{\phi}_0)^2+\tilde{V}_3(\tilde{\phi}-\tilde{\phi}_0)^3\nn\\
&+&\mathcal{O}(\tilde{\phi}^4),\label{eq:Vphi}
\ea
where we have set
{\small
\ba
\tilde{V}_0&=& \frac{1}{6}\left(4-2q_0-3\Omega_{\mathrm{m},0}\right),\\
\tilde{V}_1&=& \epsilon^{-1}\frac{8-2j_0+6q_0-9\Omega_{\mathrm{m}, 0}}{2\sqrt{6+6q_0-9\Omega_{\mathrm{m},0}}},\\
\tilde{V}_2&=& \frac{1}{8}\left(15+6q_0+\frac{4(j_0-1)^2}{\left(2+2q_0-3\Omega_{\mathrm{m}, 0}\right)^2}+\frac{4(2j_0+3q_0+s_0)}{2+2q_0-3\Omega_{\mathrm{m},0}}\right),\nn \\
\tilde{V}_3&=&\cdots,
\ea
}where we do not show the term $\tilde{V}_3$ as it is too long and complicated, but can be easily derived from the previous expressions.


We can now also calculate the effective mass of the scalar field as:
\ba
m_{\phi}^2&=&\frac{d^2 V}{d\phi^2}  \nn \\
&=&\frac{d^2 \tilde{V}}{d\phi^2} H_0^2 \nn \\
&=&2\tilde{V}_2 + 6 \tilde{V}_3 (\tilde{\phi}-\tilde{\phi}_0)+\cdots,
\ea
where the coefficients $V_2$ and $V_3$ were given earlier. Then, the second Swampland conjecture (SC2) can be written as in terms of the cosmographic parameters as
\ba
M_\textrm{pl}\frac{|V'(\phi)|}{V}&=& \frac{\tilde{V}'(\tilde{\phi})}{\sqrt{3} \tilde{V}(\tilde{\phi})}\nn \\
&=&S_0 + S_1 (\tilde{\phi}-\tilde{\phi}_0)+S_2 (\tilde{\phi}-\tilde{\phi}_0)^2+\cdots,~~~~
\ea
where the coefficients $S_0$, $S_1$ and $S_2$ are given by
\ba
S_0 &=& \frac{\tilde{V}_1}{\sqrt{3}\tilde{V}_0}, \\
S_1 &=& -\frac{\tilde{V}_1^2-2 \tilde{V}_0 \tilde{V}_2}{\sqrt{3}\tilde{V}_0^2},\\
S_2 &=&  -\frac{\tilde{V}_1^3-3\tilde{V}_0\tilde{V}_1\tilde{V}_2+3 \tilde{V}_0^2\tilde{V}_3}{\sqrt{3}\tilde{V}_0^3}.
\ea

Note that in the case of the cosmological constant model, we have that $\tilde{\phi}'(z)=0$ and   $\tilde{\phi}(z)=\tilde{\phi}_0$, which implies that $\tilde{V}(\tilde{\phi})=\tilde{V}_0=$constant, hence that $m_\phi^2=0$ and $S_0=0$ as expected. Note also that one has to take the limit to $w\rightarrow-1$ before differentiating, as the limit and the derivatives do not commute in this case.

We will present the results from the fits to the data and the cosmographic reconstructions of the conjectures in Sec.\ref{sec:results}.


\section{Data \label{sec:data}}
\subsection{The $H(z)$ data}
The Hubble expansion data is obtained by two approaches that are interdependent. First, through the clustering of galaxies or quasars, which represents a direct probe of the Hubble expansion by finding out the BAO peak in the radial direction \cite{Gaztanaga:2008xz}. Second, by the differential age method, which is connected to the redshift drift of distant objects over long time periods, normally more than a decade. This is due to the fact that in metric theories, under the assumption of the Friedmann-Robertson-Walker metric, the Hubble parameter can also be written in terms of the time derivative of the redshift as $H(z)=-\frac{1}{1+z}\frac{dz}{dt}$ \cite{Jimenez:2001gg}.

In our analysis we use the $36$ points of the compilation from Ref.~\cite{Arjona:2018jhh}, where the redshift ranges from $0.07\le z \le 2.34$ as can be seen in Table~\ref{tab:Hzdata}. By minimizing the $\chi^2$ analytically over $H_0$ we find
\ba
\chi^2_\textrm{H}&=&A-\frac{B^2}{\Gamma},\label{eq:chi2H}\\
H_0&=&\frac{B}{\Gamma},\label{eq:H0bf}
\ea
where the parameters $A$, $B$ and $\Gamma$ are defined as
\ba
A&=&\sum_i^{N_\textrm{H}}\left(\frac{H_i}{\sigma_{H_i}}\right)^2, \\
B&=&\sum_i^{N_\textrm{H}}\frac{H_i~E^\textrm{th}(z_i)}{\sigma_{H_i}^2}, \\ \Gamma&=&\sum_i^{N_\textrm{H}}\left(\frac{E^\textrm{th}(z_i)}{\sigma_{H_i}}\right)^2,
\ea
and we designate the theoretical value of the Hubble parameter as $E^\textrm{th}(z)=H^\textrm{th}(z)/H_0$ and $N_\textrm{H}=36$. This compilation may be used to measure different cosmological parameters such as the Hubble constant $H_0$, the transition redshift $z_t$, the curvature parameter $\Omega_k$ along with distance redshift data and also constrain the non-relativistic matter and DE parameters, as shown in Ref.~\cite{Yu:2017iju}.

\begin{table}[!t]
\caption{The $H(z)$ compilation used in our analysis (in units of $\textrm{km}~\textrm{s}^{-1} \textrm{Mpc}^{-1}$). This data, presented in Ref.~\cite{Arjona:2018jhh}, is partly based on those of Refs.~\cite{Moresco:2016mzx} and \cite{Guo:2015gpa}.\label{tab:Hzdata}}
\centering
\begin{tabular}{cccccccccc}
\\
\hline\hline
$z$  & $H(z)$ & $\sigma_{H}$ & Ref.   \\
\hline
$0.07$    & $69.0$   & $19.6$  & \cite{Zhang:2012mp}  \\
$0.09$    & $69.0$   & $12.0$  & \cite{STERN:2009EP} \\
$0.12$    & $68.6$   & $26.2$  & \cite{Zhang:2012mp}  \\
$0.17$    & $83.0$   & $8.0$   & \cite{STERN:2009EP}    \\
$0.179$   & $75.0$   & $4.0$   & \cite{MORESCO:2012JH}   \\
$0.199$   & $75.0$   & $5.0$   & \cite{MORESCO:2012JH}   \\
$0.2$     & $72.9$   & $29.6$  & \cite{Zhang:2012mp}   \\
$0.27$    & $77.0$   & $14.0$  & \cite{STERN:2009EP}   \\
$0.28$    & $88.8$   & $36.6$  & \cite{Zhang:2012mp}  \\
$0.35$    & $82.7$   & $8.4$   & \cite{Chuang:2012qt}   \\
$0.352$   & $83.0$   & $14.0$  & \cite{MORESCO:2012JH}   \\
$0.3802$  & $83.0$   & $13.5$  & \cite{Moresco:2016mzx}   \\
$0.4$     & $95.0$   & $17.0$  & \cite{STERN:2009EP}    \\
$0.4004$  & $77.0$   & $10.2$  & \cite{Moresco:2016mzx}   \\
$0.4247$  & $87.1$   & $11.2$  & \cite{Moresco:2016mzx}   \\
$0.44$    & $82.6$   & $7.8$   & \cite{Blake:2012pj}   \\
$0.44497$ & $92.8$   & $12.9$  & \cite{Moresco:2016mzx}   \\
$0.4783$  & $80.9$   & $9.0$   & \cite{Moresco:2016mzx}   \\
\hline\hline
\end{tabular}~~~~~~~~
\begin{tabular}{cccccccccc}
\\
\hline\hline
$z$  & $H(z)$ & $\sigma_{H}$ & Ref.   \\
\hline
$0.48$    & $97.0$   & $62.0$  & \cite{STERN:2009EP}   \\
$0.57$    & $96.8$   & $3.4$   & \cite{Anderson:2013zyy}   \\
$0.593$   & $104.0$  & $13.0$  & \cite{MORESCO:2012JH}  \\
$0.60$    & $87.9$   & $6.1$   & \cite{Blake:2012pj}   \\
$0.68$    & $92.0$   & $8.0$   & \cite{MORESCO:2012JH}    \\
$0.73$    & $97.3$   & $7.0$   & \cite{Blake:2012pj}   \\
$0.781$   & $105.0$  & $12.0$  & \cite{MORESCO:2012JH} \\
$0.875$   & $125.0$  & $17.0$  & \cite{MORESCO:2012JH} \\
$0.88$    & $90.0$   & $40.0$  & \cite{STERN:2009EP}   \\
$0.9$     & $117.0$  & $23.0$  & \cite{STERN:2009EP}   \\
$1.037$   & $154.0$  & $20.0$  & \cite{MORESCO:2012JH} \\
$1.3$     & $168.0$  & $17.0$  & \cite{STERN:2009EP}   \\
$1.363$   & $160.0$  & $33.6$  & \cite{Moresco:2015cya}  \\
$1.43$    & $177.0$  & $18.0$  & \cite{STERN:2009EP}   \\
$1.53$    & $140.0$  & $14.0$  & \cite{STERN:2009EP}  \\
$1.75$    & $202.0$  & $40.0$  & \cite{STERN:2009EP}  \\
$1.965$   & $186.5$  & $50.4$  & \cite{Moresco:2015cya}  \\
$2.34$    & $222.0$  & $7.0$   & \cite{Delubac:2014aqe}   \\
\hline\hline
\end{tabular}
\end{table}

\subsection{The growth-rate data}
We now present the growth-rate data $f\sigma_8$ that can be used to constrain the matter density parameter $\Omega_{\mathrm{m},0}$ in a model independent fashion. The data based on the  compilation provided in Table I of Ref.~\cite{Sagredo:2018ahx}, where the authors tested the internal robustness of the dataset by analyzing different subsets in the data and using Bayesian model comparison. This compilation is derived through the redshift-space distortions where it is driven the combination $f\sigma_8(a)\equiv f(a)\cdot \sigma(a)$.

The value of $f\sigma_8(a)$ can be obtained directly from the ratio of the monopole to the quadrupole of the redshift-space power spectrum, which relies on the parameter $\beta=f/b_0$, where $b_0$ is the bias and $f$ is the growth rate assuming linear theory  \cite{Percival:2008sh,Song:2008qt,Nesseris:2006er}. In Ref.~\cite{Sagredo:2018ahx} it was shown that $f\sigma_8(a)$ is independent of the bias, since the latter  cancels out from the previous expression.

The advantage of using $f\sigma_8(a)$ rather than the growth-rate $f(z)$ is that $f\sigma_8(a)$ is directly connected to the power spectrum of peculiar velocities of galaxies \cite{Nesseris:2014mea} and also it has been proven to be a good discriminator of DE models \cite{Song:2008qt}. For further details on the covariance matrix of the data and how to properly account for the Alcock-Paczynski effect, see for example Refs.~\cite{Sagredo:2018ahx}, \cite{Nesseris:2017vor} and \cite{Kazantzidis:2018rnb}. A publicly available RSD likelihood for MontePython based on the aforementioned growth-rate data compilation was initially presented in Ref.~\cite{Arjona:2020yum} and has been recently used to constrain MG models \cite{Cardona:2020ama}.

Furthermore, it was shown in Ref.~\cite{Arjona:2020kco} that the matter density parameter $\Omega_{\mathrm{m},0}$ and $\sigma_8$ can be inferred in a model independent fashion via the reconstruction of $f\sigma_8$ using the following expressions
\ba\label{eq:matter}
\Omega_{\mathrm{m},0}&=&\frac{1}{3\int_0^1 dx \frac{f\sigma_8{}(x)}{f\sigma_8{}(1)} \int_0^x dy \frac1{y} \frac{f\sigma_8{}(y)}{f\sigma_8{}(1)}} \\
\sigma_8&=&\int_0^1 \frac{f\sigma_8(x)}{x} dx \label{eq:sigma8},
\ea
which only involve integrations over the $f\sigma_8$ reconstructions and are parameter free. Note that Eqs.~(\ref{eq:matter}) and (\ref{eq:sigma8}) can be found by direct manipulations on the definition of $f\sigma_{8}$ which is defined as
\ba
f \sigma_{8}(a) & \equiv & f(a) \cdot \sigma(a) \nn\\
&=&\frac{\sigma_{8}}{\delta_{m}(1)} a \delta_{m}^{\prime}(a),
\ea
and through the differential equation that is satisfied by the growth factor $ \delta_{m}(a)$
\begin{equation}\label{eq:fs8}
\delta_{m}^{\prime \prime}(a)+\left(\frac{3}{a}+\frac{H^{\prime}(a)}{H(a)}\right) \delta_{m}^{\prime}(a)-\frac{3}{2} \frac{\Omega_{\mathrm{m}, 0}}{a^{5} H(a)^{2} / H_{0}^{2}} \delta_{m}(a)=0,
\end{equation}
where $f(a)$ is the growth rate and $\sigma(a)=\sigma_{8} \frac{\delta_{m}(a)}{\delta_{m}(1)}$ and Eq.~(\ref{eq:fs8}) equation holds assuming a homogeneous and isotropic universe with no dark energy perturbations and neglecting neutrinos.

\subsection{The $E_\textrm{g}$ data}
The perturbed flat Friedmann-Lemaitre-Robertson-Walker (FLRW) metric, in the conformal Newtonian gauge, is defined as
\be
ds^2=-(1+2\Psi)dt^2 +a(t)^2(1-2\Phi)dx^2,
\ee
where the terms $\Psi$ and $\Phi$ represent the two scalar gravitational potentials and $a$ is the scale factor. The ratio of the gravitational potentials define what is known as the gravitational slip $\eta=\frac{\Phi}{\Psi}$ which has to be unity in GR. These potentials must obey the two Poisson equations in Fourier space:
\ba
-\frac{k^2}{a^2}(\Phi+\Psi)&=&4 \pi G_\textrm{N} \Sigma(k,a) \rho_m \delta_m, \\
-\frac{k^2}{a^2}\Psi&=&4 \pi G_\textrm{N} \mu(k,a) \rho_m \delta_m,
\ea
where $G_\textrm{N}$ is the bare Newton's constant and $\Sigma$ and $\mu$ parameterize deviations in GR, which is recovered for $\Sigma=2$ and $\mu=1$.

Recently, the $E_\textrm{g}$ statistic was used to test the aforementioned relations, with the hope of being model independent at the linear order \cite{Zhang:2007nk,Reyes:2010tr}. This $E_\textrm{g}$ test can also be written as the expectation value of the ratio of lensing and galaxy clustering observables at a scale $k$ as follows
\be
E_\textrm{g}=\left\langle \frac{a \nabla^2(\Psi+\Phi)}{3 H_0^2 f \delta_m}\right\rangle.
\ee
To perform a model independent reconstruction on the gravitational slip $\eta$ we reconstruct two quantities through the $E_\textrm{g}$ and $f\sigma_8$ data. First, the quantity $P_2(z)$ defined as $P_2=\frac{\Omega_\textrm{m,0}\Sigma}{f}$ which depends on the lensing potential and the growth rate. In GR this simplifies to $P_2=\frac{2\Omega_\textrm{m,0}}{f}$, which for GR we then have $E_\textrm{g}=\frac{\Omega_\textrm{m,0}}{f}$. In general, $E_\textrm{g}$ can be connected to the $P_2$ statistic of Ref.~\cite{Pinho:2018unz} as $P_2=2 E_\textrm{g}$. Second, the quantity $P_3$ defined as $P_3=\frac{\left(f\sigma_8(z)\right)'}{f\sigma_8(z)}$, where the prime is the derivative with respect to $\ln a$. Then $\eta$ can be defined as \cite{Pinho:2018unz}
\be
\eta(z)=\frac{3P_2(z)(1+z)^3}{2E(z)^2\left(P_3(z)+2+\frac{E'(z)}{E(z)}\right)}-1, \label{eq:etade}
\ee
where $E(z)\equiv H(z)/H_0$. For completeness the data points used in our analysis are given in Table~\ref{tab:Eg}.

\begin{table}[!t]
\caption{The $E_\textrm{g}$ data used in this analysis as compiled by Refs.~\cite{Pinho:2018unz} and \cite{Skara:2019usd}. Notice that some of the points in the references mentioned before were duplicates as they come from the same surveys, even though with combinations of different external probes, so we use only one of the measurements to avoid strong correlations. Here we only display the points we used in the analysis.\label{tab:Eg}}
\begin{centering}
\begin{tabular}{cccc}
 $z$ & $E_\textrm{g}$  &  $\sigma_{E_\textrm{g}}$ & \\ \hline
0.267 &	0.43 &	0.13 & \\
0.270 & 0.40 &  0.05 & \\
0.305 &	0.27 &	0.08 & \\
0.320 &	0.40 &	0.09 & \\
0.554 &	0.26 &	0.07 & \\
0.570 &	0.30 &	0.07 & \\
0.600 &	0.16 &	0.09 & \\
0.860 &	0.09 &	0.07 &
\end{tabular}
\par
\end{centering}
\end{table}
\section{Genetic Algorithms \label{sec:GA}}
The Genetic Algorithms (GA) are a group of machine learning (ML) techniques that are designed to perform non-parametric reconstruction of data and are constructed on the concept of grammatical evolution, conveyed by the genetic operations of crossover and mutation. For an in-depth discussion on the GA and several applications to cosmology see Refs.~\cite{Bogdanos:2009ib,Nesseris:2010ep, Nesseris:2012tt,Nesseris:2013bia,Sapone:2014nna,Arjona:2020doi,Arjona:2020kco,Arjona:2019fwb,Arjona:2020axn}. The GA emulate the idea of evolution by the application of natural selection; a group of individuals evolves over time under the effect of the stochastic operators of mutation, i.e a random modification in an individual, and crossover, i.e. the merger of two or more different individuals, producing two new individuals (children). Every individual's ``reproductive success'', usually quantified through a $\chi^2$ statistic, is taken to be proportional to its fitness, measuring how precisely each individual of the population fits the data.

In what follows, we will outline how to reconstruct the Hubble parameter $H(z)$ from the Hubble expansion history $H(z)$ data, $f\sigma_8(z)$ from the growth-rate data derived via the redshift-space distortions
(RSD) and $P_2(z)$ from the Eg data. The reconstruction of $H(z)$, along with that of $H_0$, is needed for Eqs.~(\ref{eq:Vx}) and (\ref{eq:phix}), while our reconstruction for $f\sigma_8(z)$ is used to infer model independently the value of $\Omega_{\mathrm{m},0}$ by means of Eq.~(\ref{eq:matter}). Our reconstruction of $P_2(z)$ from the Eg data is used at a latter stage in Sec.~\ref{sec:diagrams}.

The outline to perform the reconstructions proceeds as follows. An initial population of functions is randomly selected so that every member of the population holds initial guesses for $H(z)$, $f\sigma_8(z)$ and $P_2(z)$. We also impose reasonable physical priors, e.g. the Hubble parameter today is given by the Hubble constant $H(z = 0) = H_0$, which then allows us to estimate $H_0$ directly from the $H(z)$ data. For the $f\sigma_8(z)$ reconstruction we assume that the Universe at early times went through a phase of matter domination $(z\simeq100)$, then the linear growth acts as $\delta_m(a)\simeq a$ at high redshifts. However we make no assumption of a DE model or on the curvature of the Universe.

Next, the fitness of each member is computed through a $\chi^2$ statistic, using as input the $H(z)$, growth and Eg data. Afterwards, the mutation and crossover operators are applied stochastically to the best-fitting functions in each generation, selected via the tournament selection method, see for more details \cite{Bogdanos:2009ib}. This process is then repeated with different random seeds thousands of times in order to ensure convergence and not to bias the results due to a specific choice of the random seed. After the GA code has converged, the final output is a reconstruction of $H(z)$, $f\sigma_8(z)$ and $P_2(z)$.

For the estimation of the errors on the reconstructed functions we implement an analytical approach developed by Refs.~\cite{Nesseris:2012tt,Nesseris:2013bia}, where the errors are derived via a path integral over the whole functional space that can be scanned by the GA. The GA path integral approach has been compared with bootstrap Monte-Carlo error estimates \cite{Nesseris:2012tt} finding excellent agreement with both approaches.

To sum up, with the GA we can reconstruct any cosmological function, for example the $H(z)$, $f\sigma_8(z)$ and $P_2(z)$ that we consider here, by applying the algorithm to any dataset of interest. There are no requirements on the specific cosmological model or assumptions on DE, hence our results are model independent. Besides executing a large number of GA runs with different random seed numbers, in order to avoid spurious reconstructions and overfitting we have imposed that all reconstructed functions, as well as their derivatives, are continuous in the range of redshifts we consider.
\section{Results \label{sec:results}}
\subsection{Genetic Algorithm reconstructions}
In this section we will now discuss our ML fits to the data and the corresponding reconstructions of the Swampland conjectures. First, we show in Table~\ref{tab:GAchi2} the best-fit $\chi^2$ for the GA functions for the \lcdm model. As can be seen, in all cases the GA out-performs the \lcdm model in terms of the best-fit $\chi^2$.

\begin{figure*}[!t]
\centering
\hspace*{-4mm}
\includegraphics[width = 0.45\textwidth]{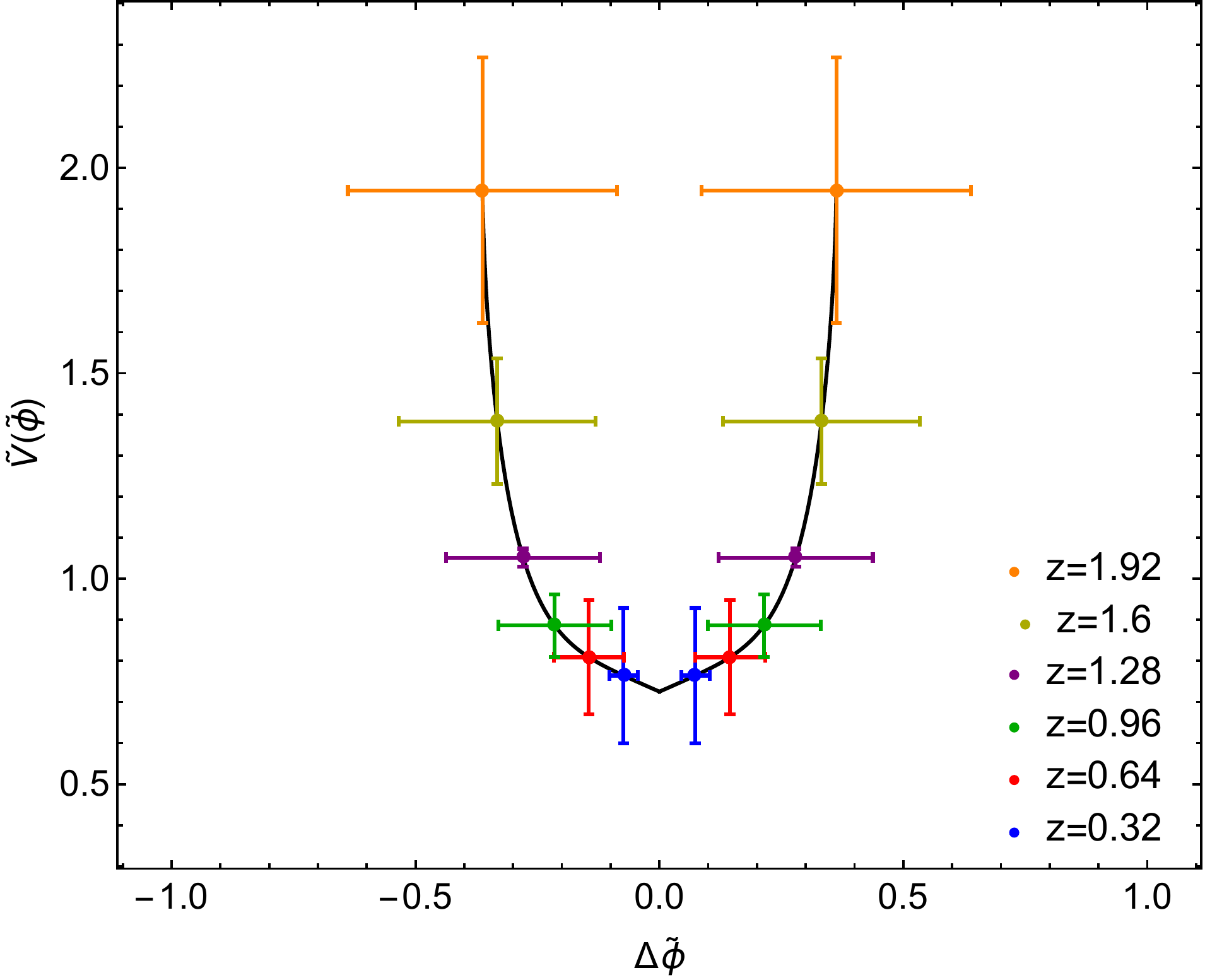}
\includegraphics[width = 0.48\textwidth]{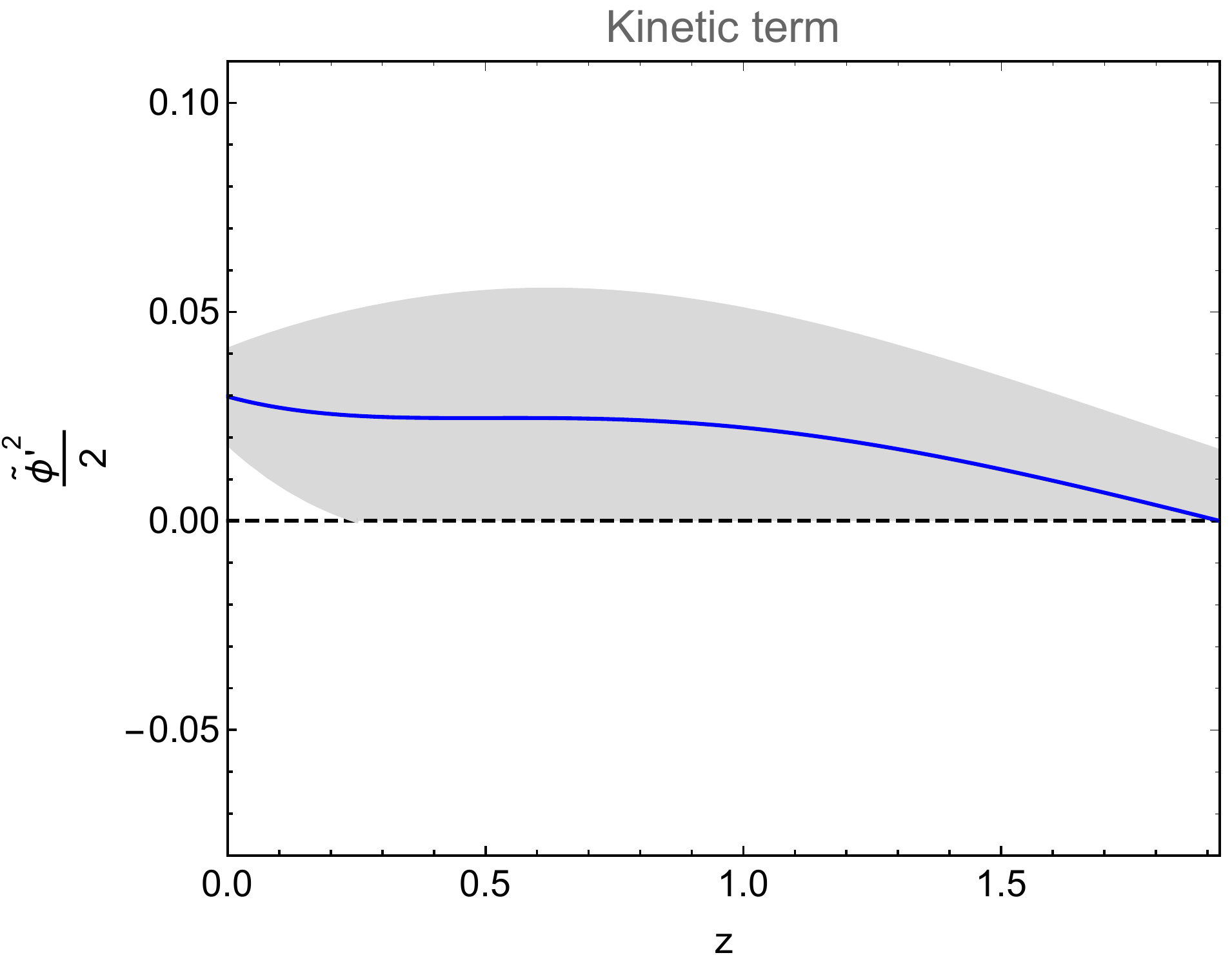}
\caption{The GA reconstruction of the scalar field potential (left) and the kinetic term (right) for the redshift range $z \in [0,1.92]$. The left panel suggests a parabolic shape of the potential, thus pointing toward some deviations from the \lcdm model since for the latter the potential should be constant. The black solid line corresponds with the GA best-fit, the different colors represent the errors for our reconstructions at different redshifts and that $\Delta \tilde{\phi}=\tilde{\phi}(z)-\tilde{\phi}_0$. In the right panel we see that the reconstructed kinetic term is positive from $z=0$ to $z~\sim 1.9$, hence our reconstructions of the Swampland conjectures within that redshift range should be free from ghosts and instabilities. The blue solid line corresponds to the GA best-fit and the grey region to the $1\sigma$ errors. Note that in the right panel we truncate the error in the unphysical region where the kinetic term changes sign. \label{fig:pot_errkin}}
\vspace*{-1mm}
\end{figure*}

Then, in the left panel of Fig.~\ref{fig:pot_errkin} we present our GA reconstruction of the scalar field potential as a function of the scalar field for the redshift range $z \in [0,1.92]$. We can see a parabolic shape of the potential, thus pointing toward some deviations from the \lcdm model since for the latter the potential should be flat. The black solid line corresponds with the GA best-fit and the different colours represent the errors for our reconstructions at different redshifts. In the right panel of Fig.~\ref{fig:pot_errkin} we see that the reconstructed kinetic term is positive from $z=0$ to $z~\sim 1.9$, hence our reconstructions of the Swampland conjectures within that redshift range should be free from ghosts and instabilities in this redshift range. The blue solid line corresponds to the GA best-fit and the grey region to the $1\sigma$ errors.

Note that the fact that at some redshift the kinetic term of the quintessence scalar field is negative, is actually a common issue of such reconstruction methods. This happens not only for quintessence models, but even for scalar tensor theories, where one naively may expect that due to the extra degree of freedom this would not happen\footnote{See for example Fig.~5 in Ref.~\cite{Nesseris:2006jc}, where a similar phenomenon is also observed.}. This simply means that at some redshift the reconstruction breaks down, as obviously the kinetic term has to be both real and positive. However, due to the larger errors this is not a big problem per se, it just limits our ability to use this particular model at all redshifts.

\begin{figure*}[!t]
\centering
\hspace*{-4mm}
\includegraphics[width = 0.48\textwidth]{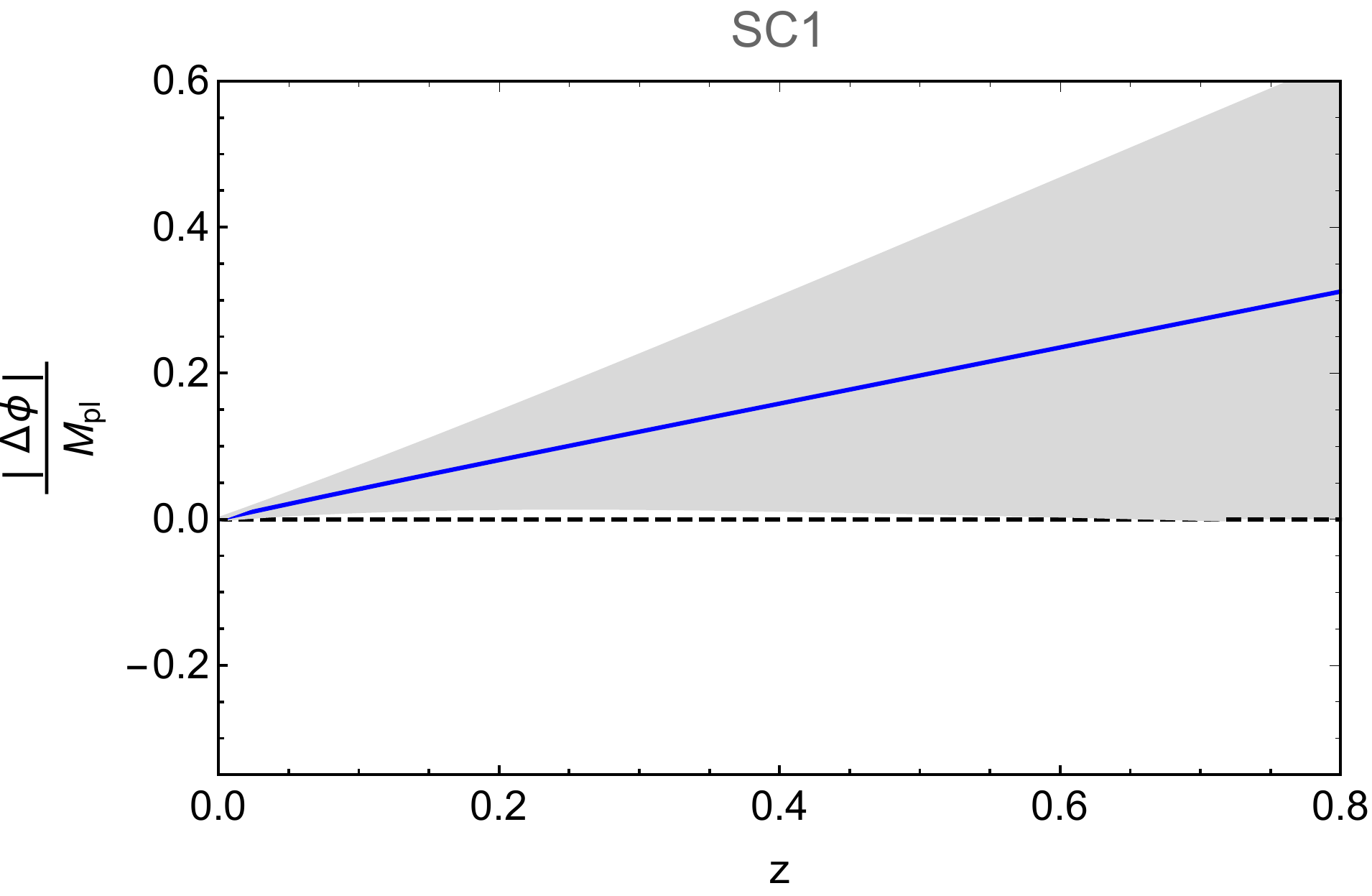}
\includegraphics[width = 0.48\textwidth]{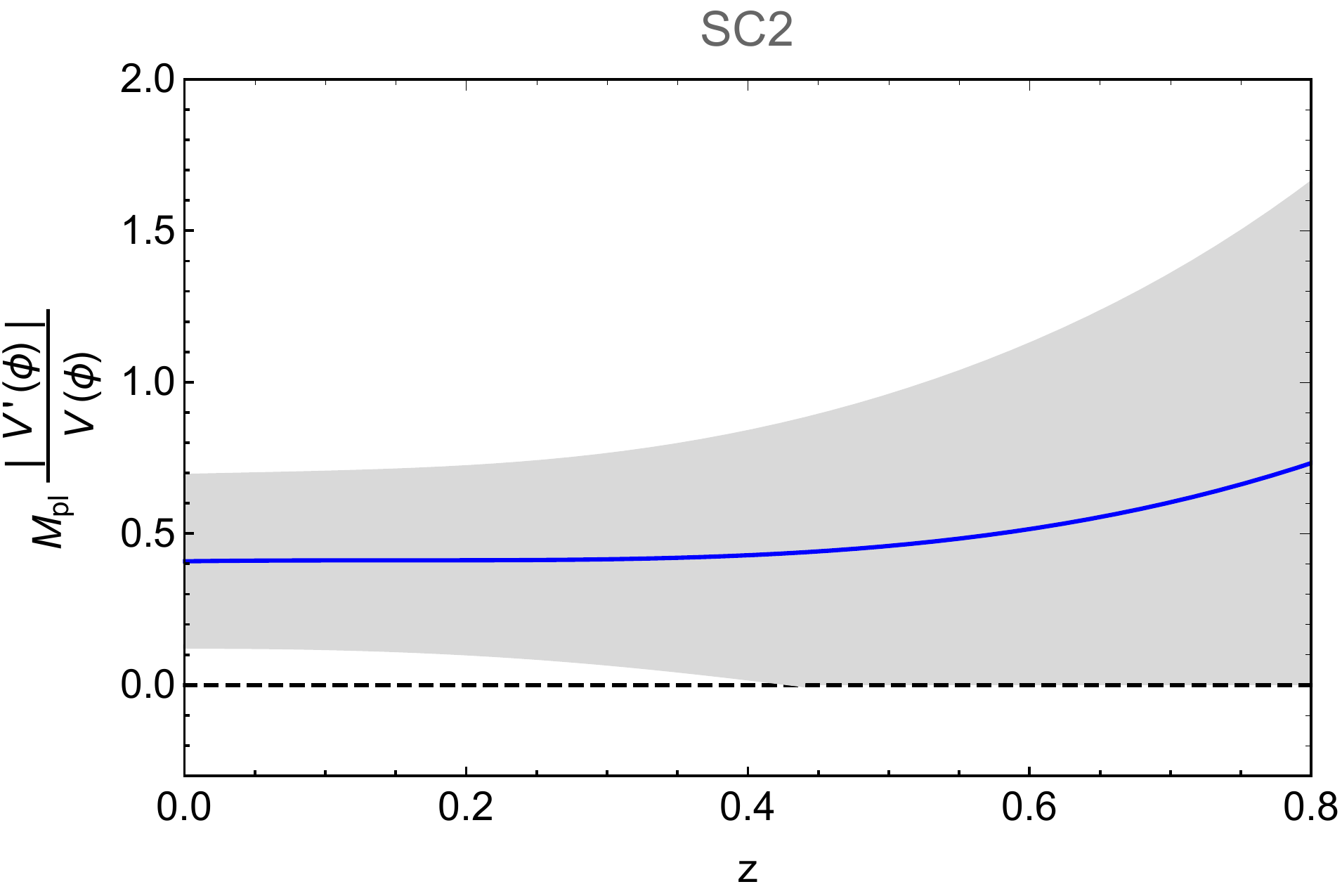}
\caption{The GA reconstruction of the first and second Swampland conjectures on the left and right panels respectively. As can be seen, from $z=0.8$ onward the errors become so large that we can only draw some conclusions at low redshifts with the current available data. Both reconstructions are consistent with the conjectures and the \lcdm model at the $1\sigma$ level. Notice that $\Delta \phi=\phi(z)-\phi_0$ and that we truncate the error in the unphysical region where the quantities would change sign (as we take the absolute values). \label{fig:pot_err}}
\vspace*{-1mm}
\end{figure*}

In Fig.~\ref{fig:pot_err} we present our GA reconstructions of the first and second Swampland conjectures on the left and right panels respectively. In both cases the blue solid line and the grey region corresponds to the GA best-fit and the $1\sigma$ error respectively. As can be seen, from $z=0.8$ onward the errors become so large that we can only draw some conclusions at low redshifts with the current available data. In specific we can see how both reconstructions are consistent with the conjectures, being both order unity, although at the same time being consistent with the \lcdm model at the $1\sigma$ level.

\begin{table}[]
    \centering
     \begin{tabular}{cccc}
Model & $H(z)$ & $f \sigma_{8}$ & $P_{2}$ \\
\hline$\chi_{\Lambda \mathrm{CDM}}^{2}$ & 19.476 &   12.238 & 10.516 \\
\hline$\chi_{G A}^{2}$ & 17.670 & 12.220 & 5.422 \\
\hline
\end{tabular}
\caption{The $\chi^2$ for \lcdm and GA using the growth $f\sigma_8$, the Hubble rate $H(z)$ and the $E_g$ statistics.}
\label{tab:GAchi2}
\end{table}

\subsection{Cosmographic reconstructions}
We now present the cosmographic reconstructions of the potentials and the first and second Swampland conjectures. We find that the standard quintessence reconstruction via cosmography suffers from two issues: first, it is only valid when the square of the kinetic term given via Eq.~\eqref{eq:phix0} is positive. Second, a value of the matter density $\Omega_{\mathrm{m},0}$ from an external source is required, as the $H(z)$ cannot provide it since Eq.~\eqref{eq:cosmoH} only depends on the cosmographic parameters and not at all on $\Omega_{\mathrm{m},0}$.

We address the first issue by performing the reconstruction only in the redshift range where the kinetic term is positive, while for the value of the matter density parameter we will assume in what follows the Planck prior $\Omega_m=0.315 \pm0.007$ \cite{Aghanim:2018eyx}.

\begin{table*}[]
    \centering
    \begin{tabular}{c|cccccccc}
Order/param.  & $q_0$   & $j_0$  & $s_0$   & $l_0$   & $h$ & $\chi^2$ & $|\Delta \textrm{AIC}_c|$& $|\Delta \textrm{BIC}|$\\\hline
2nd   & $0.144\pm0.093$ &$-$ & $-$ & $-$ & $0.586\pm0.023$ & $24.772$ & $5.302$& $5.302$\\
3rd   & $-0.347\pm0.195$ &$0.432\pm0.247$ & $-$ & $-$ & $0.662\pm0.040$ & $19.172$& $2.088$& $3.286$\\
4th   & $-0.851\pm0.373$ &$2.222\pm1.412$ & $2.326\pm4.029$ & $-$ & $0.718\pm0.057$ & $17.301$& $2.758$& $4.998$\\
5th   & $-0.824\pm0.727$ &$2.133\pm3.951$ & $2.091\pm12.525$ & $9.661\pm34.294$ & $0.714\pm0.075$ & $17.304$& $5.470$& $8.585$\\
    \end{tabular}
\caption{The best fit parameters of the cosmographic expansions. For comparison the \lcdm model has best-fit parameters $(\Omega_{\mathrm{m},0},h)=(0.259\pm0.029,0.704\pm0.023)$ with $\chi^2=19.470$, while the GA has $\chi^2=17.670$. The values of the differences for the AIC and BIC, are given with respect to those of the \lcdm model.}
    \label{tab:cosmoparams}
\end{table*}

\begin{figure*}[!t]
\centering
\hspace*{-4mm}
\includegraphics[width = 0.49\textwidth]{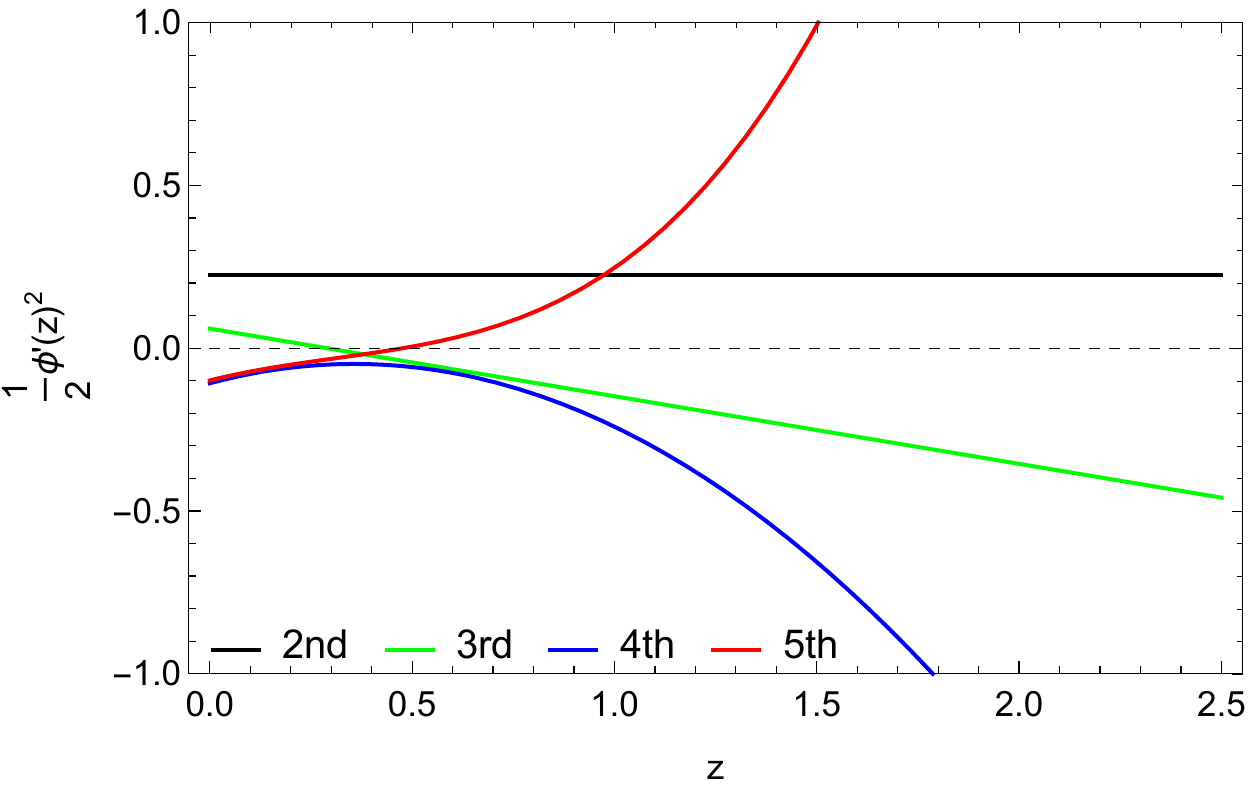}
\includegraphics[width = 0.48\textwidth]{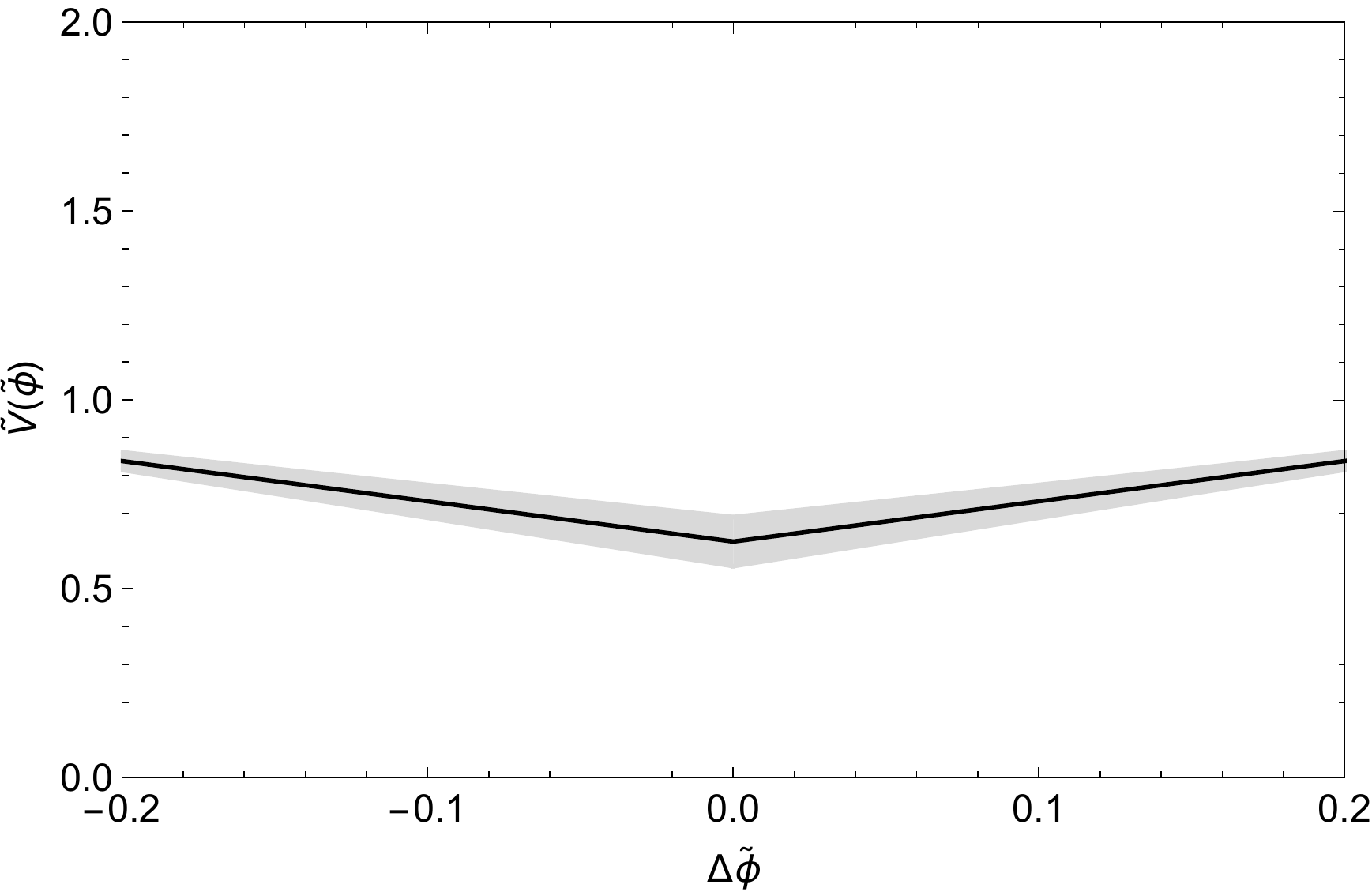}
\caption{Left: Reconstruction of the kinetic term of the scalar field for various orders of the cosmographic expansion for the Planck prior $\Omega_m=0.315 \pm0.007$ \cite{Aghanim:2018eyx}. As can be seen the third order expansion is positive in the range $z\in[0,0.3]$, while in the other cases the reconstruction breaks down as the field is complex. Right: the third order cosmographic reconstruction on the scalar field potential, where  $\Delta \tilde{\phi}=\tilde{\phi}(z)-\tilde{\phi}_0$. We only plot the quantities in the regions where the scalar field kinetic term is positive, which in this case is in the range $z\in[0,0.3]$. As can be seen, in the aforementioned redshift range the cosmographic and the GA reconstructions are consistent.\label{fig:pot_errcosm0}}
\vspace*{-1mm}
\end{figure*}

\begin{figure*}[!t]
\centering
\hspace*{-4mm}
\includegraphics[width = 0.49\textwidth]{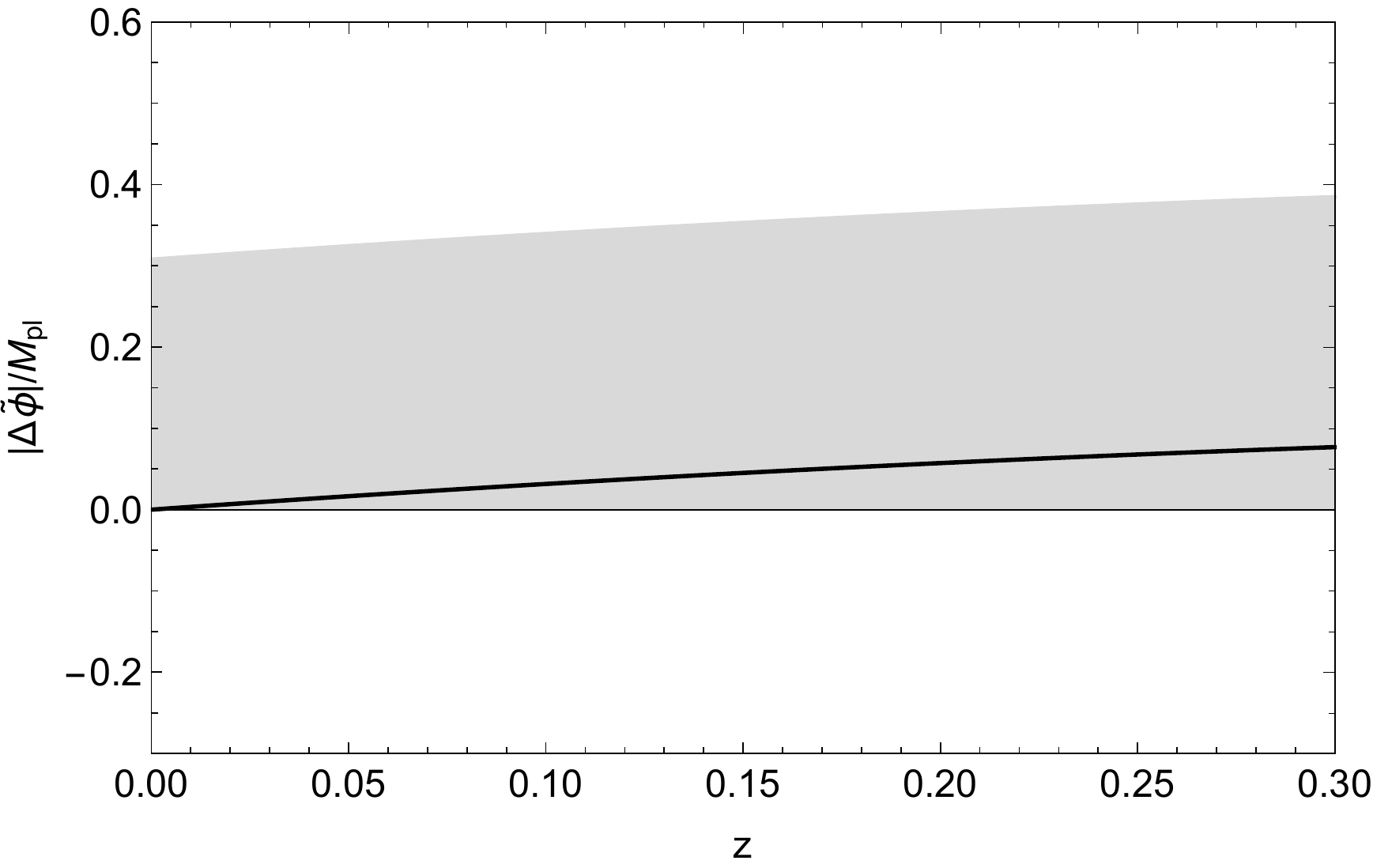}
\includegraphics[width = 0.49\textwidth]{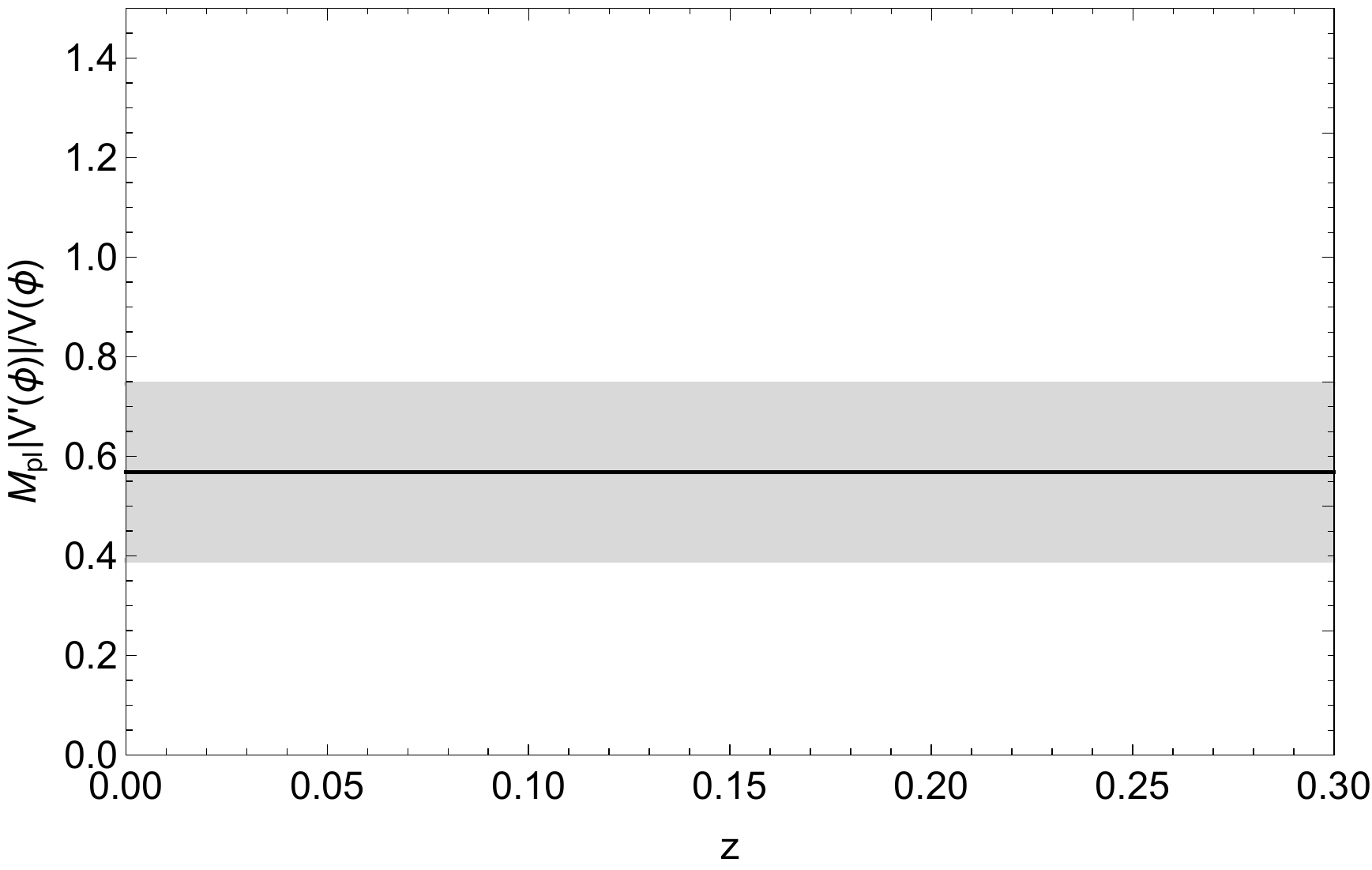}
\caption{The third order cosmographic reconstruction on the scalar field potential (left) and the Swampland conjectures SC1 (center) and SC2 (right), where $\Delta \tilde{\phi}=\tilde{\phi}(z)-\tilde{\phi}_0$. We only plot the quantities in the regions where the scalar field kinetic term is positive, which in this case is in the range $z\in[0,0.3]$. As can be seen, in the aforementioned redshift range the cosmographic and the GA reconstructions of the Swampland conjectures are consistent. Note that we truncate the error in the unphysical region where the quantities would change sign (as we take the absolute values).\label{fig:pot_errcosm}}
\vspace*{-1mm}
\end{figure*}

Next, in Table~\ref{tab:cosmoparams} we show the best-fit parameters of the cosmographic parameters for the $H(z)$ data and their $\chi^2$ values for various orders of the series expansion. We also give the values of the corrected AIC and BIC tools, which are defined as \cite{Liddle:2007fy}:
\ba
\textrm{AIC}&=& \chi^2_\textrm{min}+2k,\\
\textrm{AIC}_c&=& \textrm{AIC}+\frac{2k(k+1)}{N-k-1},\\
\textrm{BIC} &=& \chi^2_\textrm{min}+k \ln N,
\ea
where $k$ is the number of free parameters and $N$ the number of data points, which for the Hubble data is equal to $N=36$, while $\chi^2_\textrm{min}$ is the value of the $\chi^2$ at the minimum.

In Table~\ref{tab:cosmoparams} the values of the differences of the AIC${}_c$ and BIC are given with respect to those of the \lcdm model. In either case, the condition $\Delta \textrm{AIC}_c\le 2$ implies the consistency between the models, while the inequalities $2<\Delta \textrm{AIC}_c\le4$ and $4<\Delta \textrm{AIC}_c\le 7$ indicate a mild and  positive evidence against the model with higher value of AIC. On the other hand, when $\Delta \textrm{AIC}_c \ge 10$ suggests strong evidence. The same applies to the BIC test \cite{Nesseris:2012cq}. As can be seen, in all cases the \lcdm model seems to be favored by the data, which is also consistent with previous analyses \cite{Basilakos:2018arq}. Note that we cannot apply either of the criteria to the GA as it is a non-parametric approach.

Going back to the scalar field, the aforementioned issue with the kinetic term is also shown in Fig.~\ref{fig:pot_errcosm0} where in the left panel we show a reconstruction of the kinetic term of the scalar field for various orders of the cosmographic expansion for the Planck prior $\Omega_m=0.315 \pm0.007$ \cite{Aghanim:2018eyx}. As can be seen the third order expansion is positive in the range $z\in[0,0.3]$, while in the other cases the reconstruction breaks down as the field is complex. Thus, out of all cases only the second and third order expansions have a positive kinetic term at small redshifts and from these two, only the third order expansion has non-trivial phenomenology, hence we focus on this in what follows.

In the right panel of Fig.~\ref{fig:pot_errcosm0} we show the third order cosmographic reconstruction on the scalar field potential. We note that we only plot the quantities in the regions where the scalar field kinetic term is positive, which in this case is in the range $z\in[0,0.3]$. As can be seen, in the aforementioned redshift range the cosmographic and the GA reconstructions are consistent.

Finally, we show the results for the Swampland conjecture reconstructions in Fig.~\ref{fig:pot_errcosm}. As can be seen  both cases are in good agreement in the range $z\in [0,0.3]$ within the errors with the corresponding GA reconstructions. However, given that the reconstructions are limited in range, this also significantly reduces their appeal.


\section{Phase Diagrams \label{sec:diagrams}}
In MG theories it is frequently assumed that the background level behaves as the \lcdm model and its perturbations evolve differently, see for example Ref.~\cite{Troster:2020kai}. This happens in well known DE and MG models such as the so-called designer $f(R)$ models \cite{Multamaki:2005zs,delaCruzDombriz:2006fj,Pogosian:2007sw,Nesseris:2013fca,Arjona:2018jhh} or the designer Horndeski family of models (HDES) \cite{Arjona:2019rfn} whose background is exactly that of the \lcdm model. However, in general, MG models can have departures from \lcdm both at the background and at the perturbation level, as is for example the case for the Hu-Sawicki $f(R)$ gravity model \cite{Hu:2007nk} or the Kinetic Gravity Braiding (KGB) theory \cite{Kimura:2010di}.

In this section we search for potential deviations from \lcdm at the background and at the perturbation level by applying our Machine Learning reconstructions to cosmological observations. In specific we use $H(z)$ expansion rate data, the growth-rate $f\sigma_8(z)$ compilation and the Eg statistics to present two phase diagrams, $H-f\sigma_8$ and $\eta-f\sigma_8$, where $\eta$ is a parameter which defines the departure of gravity from GR and can be explored by gravitational lensing.

The anisotropic stress $\eta$ is conventionally modeled through the parameter $\eta=\frac{\Phi}{\Psi}$, where $\Phi$ and $\Psi$ are the Newtonian potentials and are considered  equal in GR in the absence of anisotropic
stresses from other sources like neutrinos. Then,
any departure of $\eta$ from unity would hint to modified gravity or if neglected, it could bias the cosmological parameters inferred from the data \cite{Arjona:2019zqg}.

\begin{figure*}[!t]
\centering
\hspace*{-4mm}
\includegraphics[width = 0.48\textwidth]{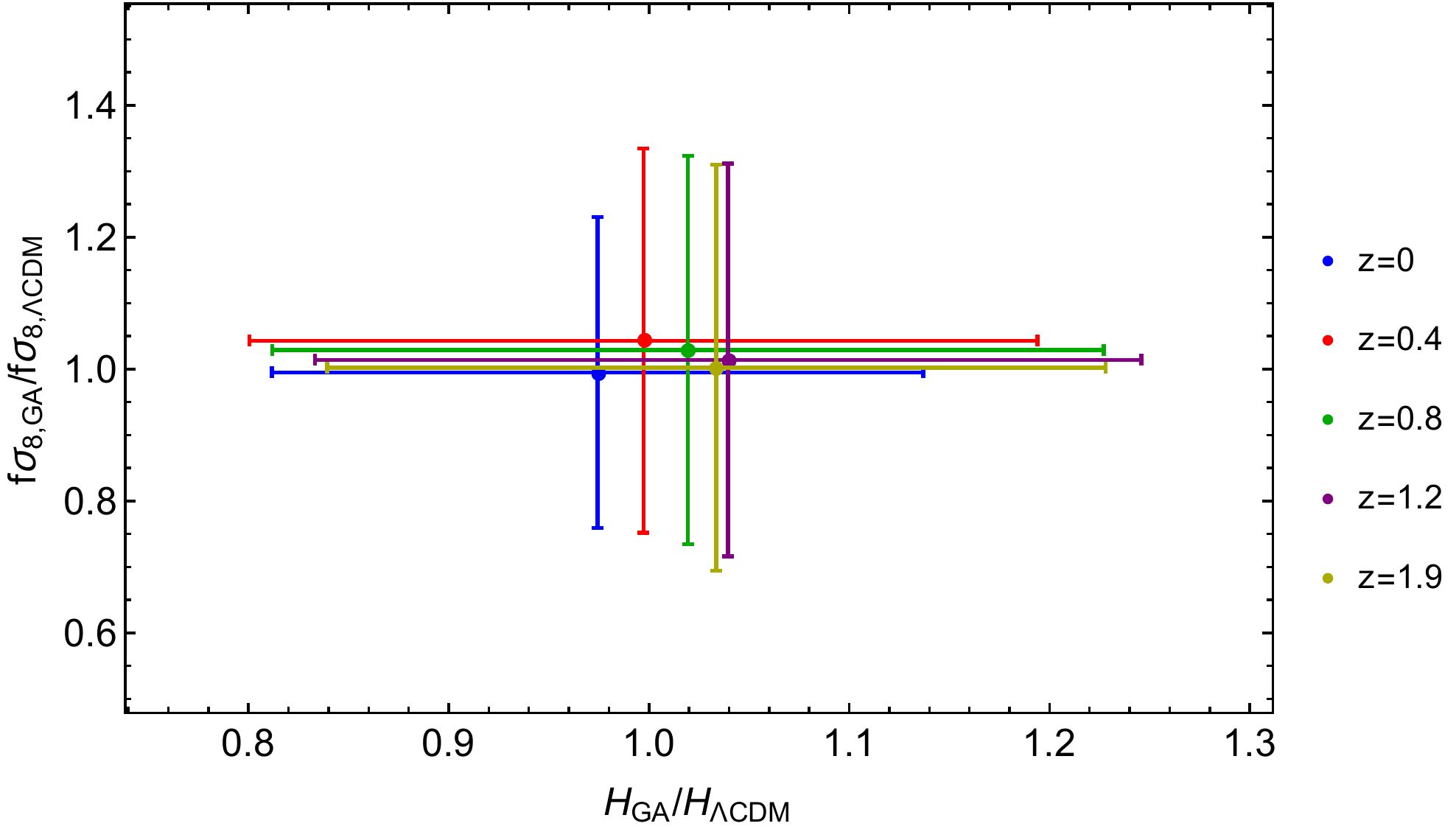}
\includegraphics[width = 0.48\textwidth]{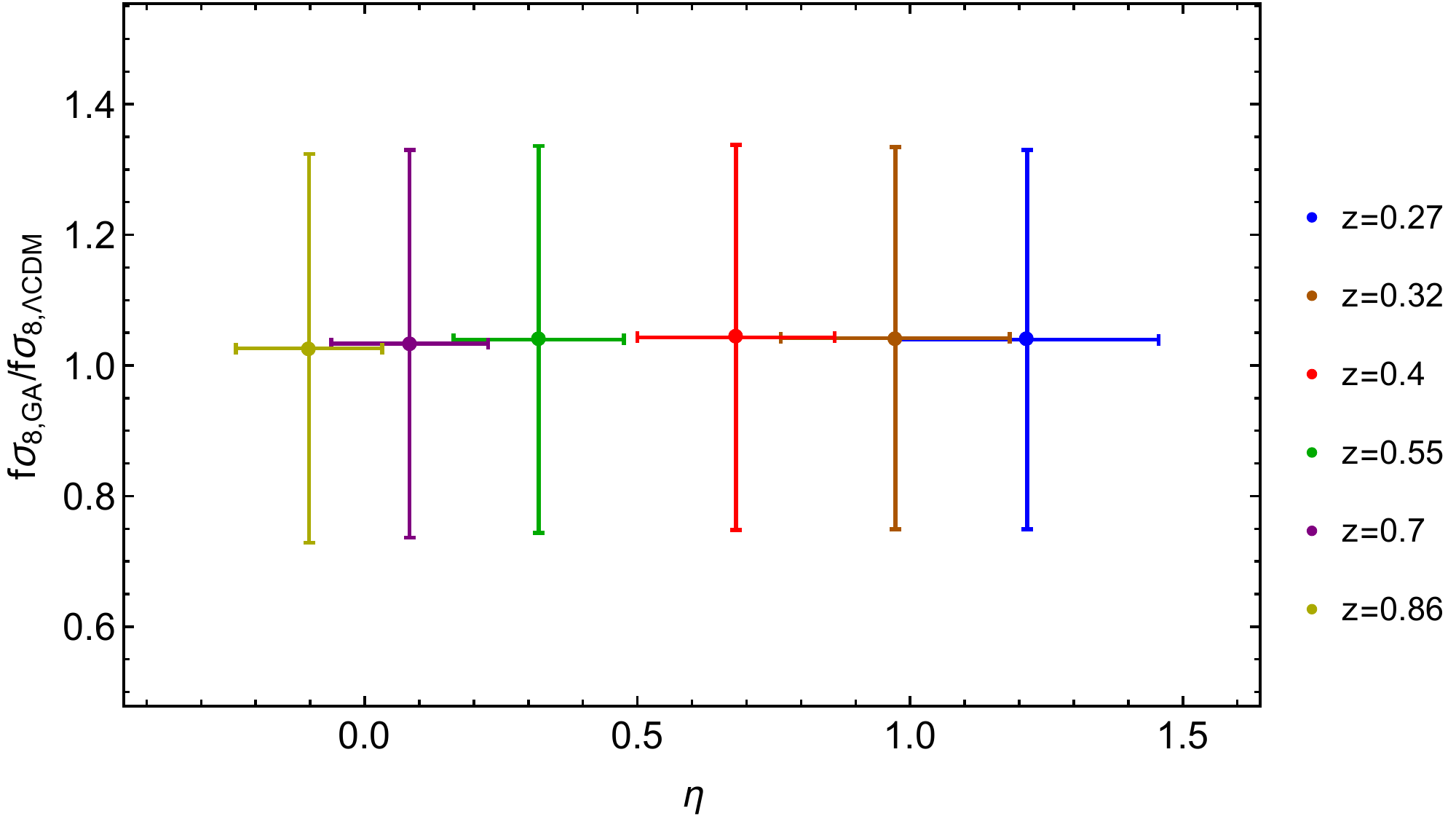}
\caption{Left: $H-f\sigma_8$ diagram for the redshift range of $z\in \left[0,1.9\right]$ using the $H(z)$ and $f\sigma_8$ data. The theoretical prediction of the \lcdm model would correspond to a point at $(1,1)$, something which the GA best-fit reconstructions confirm within the errors. The different colors represent our reconstructions for several redshifts with their respective errors. Right: $\eta-f\sigma_8$ diagram for the redshift range of $z\in \left[0.27,0.86\right]$ using the $E_g$ and $f\sigma_8$ data. The theoretical prediction of the \lcdm model would correspond to a point at $(1,1)$. The different colours represent our reconstructions for several redshifts with their respective errors. With our GA best-fit reconstructions we find a $\sim 2\sigma$ deviation of the anisotropic stress from unity at $z\sim 0.3$ and a $\sim 4\sigma$ at $z\sim 0.9$.\label{fig:fs8-H}}
\vspace*{-1mm}
\end{figure*}

With our GA reconstructions we demonstrate how deviations from the \lcdm model appears in the $\eta-f\sigma_8$ diagram. In specific, in the left panel of Fig.~\ref{fig:fs8-H}, we plot the $H-f\sigma_8$ diagram for the redshift range of $z\in \left[0,1.9\right]$ using the $H(z)$ and $f\sigma_8$ data. The theoretical prediction of the \lcdm model would correspond to a point at $(1,1)$, something which the GA best-fit reconstructions confirm within the errors. The different colors represent our reconstructions for several redshifts with their respective errors.

In the right panel of Fig.~\ref{fig:fs8-H} we have the $\eta-f\sigma_8$ diagram for the redshift range of $z\in \left[0.27,0.86\right]$ using the $E_g$ and $f\sigma_8$ data. The theoretical prediction of the \lcdm model would correspond to a point at $(1,1)$. The different colors represent our reconstructions for several redshifts with their respective errors. With our GA best-fit reconstructions we find a $\sim 2\sigma$ deviation of the anisotropic stress from unity at $z\sim 0.3$ and a $\sim 4\sigma$ at $z\sim 0.9$, thus hinting toward some deviations from GR, something which was also seen in Ref.~\cite{Arjona:2020kco}.

\begin{figure}[!t]
\centering
\hspace*{-4mm}
\includegraphics[width = 0.48\textwidth]{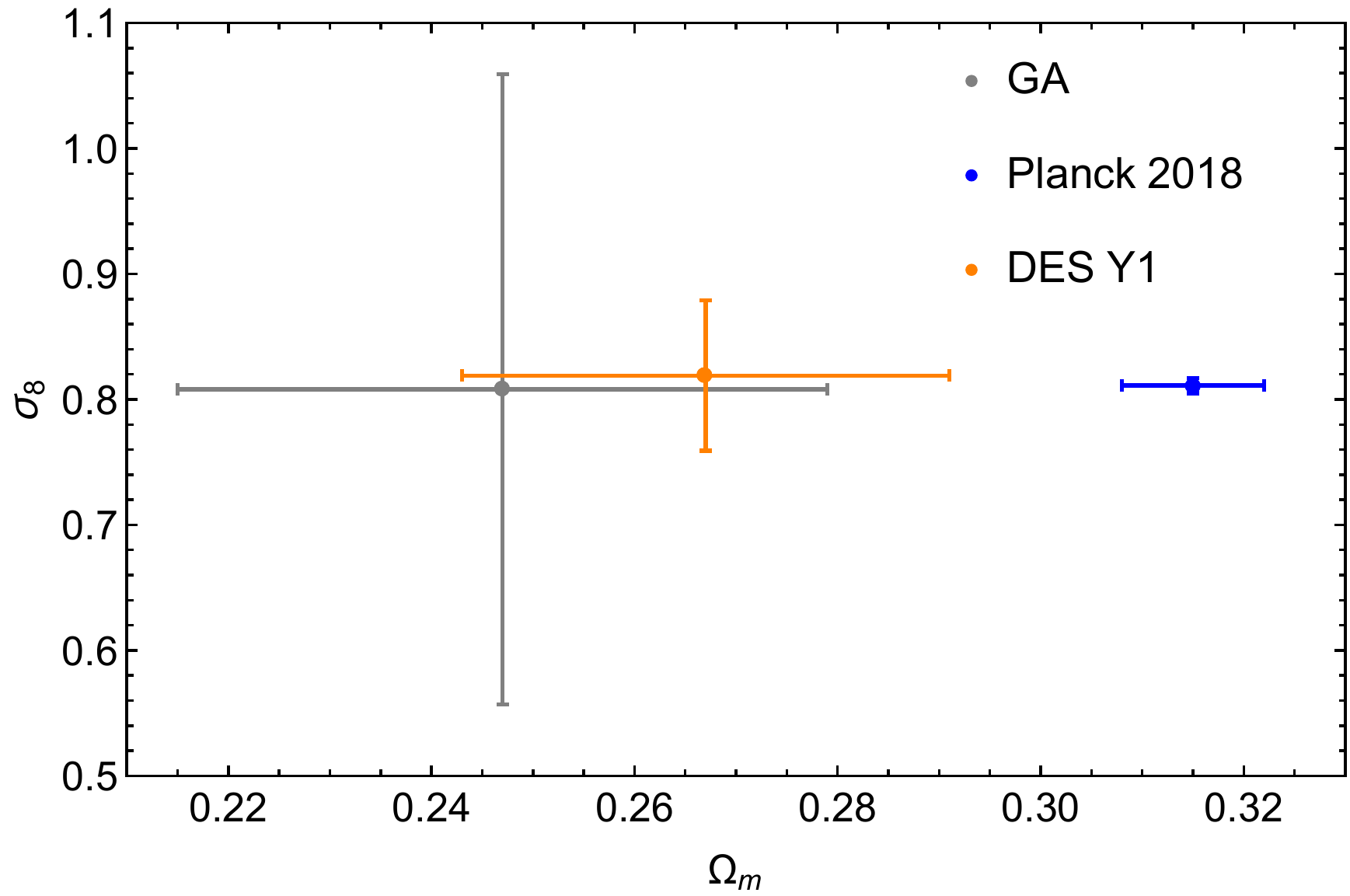}
\caption{Plot for $\sigma_8$ against $\Omega_m$ from our ML reconstructions on the growth rate $f\sigma_8$ data (gray point), the Planck 2018 TT,TE,EE+lowE+lensing values (blue point) and the DES Y1 2018 results (orange point). \label{fig:S8tension}}
\vspace*{-1mm}
\end{figure}

These deviations could be due to new physics or for the presence of unaccounted systematics and perhaps non-negligible radiative processes, as could be the case for the $E_g$ data. The diagrams presented will be useful for future large-scale structure observations to rule out some of the modified gravity models.

Finally, we also report our results from our ML reconstruction through Eqs.~(\ref{eq:matter}) and (\ref{eq:sigma8}) for the quantity $S_8$, quantified as $S_8=\sigma_8 \sqrt{\Omega_m / 0.3}$. This parameter is in tension between Planck and cosmic shear measurements such as Kids-1000 and DES Y1 above the 2$\sigma$ level, where the latter have a preference for a lower value, see for example Fig.~1 of \cite{DiValentino:2020vvd} where the tension can be visualized in the $\sigma_8 - \Omega_{\mathrm{m},0}$ diagram. With the GA we find $S_8=0.733\pm 0.275$ and in Fig.~\ref{fig:S8tension} we also provide a similar diagram, being our results more in agreement with DES Y1. Overall, these results are consistent with Ref.~\cite{DiValentino:2020vvd} and the analysis of \cite{Muir:2020puy} from the DES Y1 release, where the authors split the matter density into a ``geometric" and ``growth" $\Omega_{\mathrm{m},0}$. They are in mild tension with the Planck 2018 results though, something that has been noted repeatedly in the literature, see for example Ref.~\cite{Nesseris:2017vor} and references there in. Note that the errors with the GA approach are considerably larger than those from DES Y1 and the Planck 2018 results being one of the reasons the fact that with our reconstructions we are not making assumptions on flatness or a dark energy model, hence more agnostic.

\section{Conclusions \label{sec:conclusions}}
Through a plethora of cosmological probes it has been established that the Universe is expanding in an accelerated way, being one of the biggest findings in modern cosmology and implying that $70\%$ of the mass-energy density is made of an unknown content, usually referred to as dark energy (DE), spreading uniformly over the Universe. The cosmological constant $\Lambda$ represents one of the simplest contenders of DE, however some of it caveats is that the observed energy scale of DE is much smaller than the energy scale predicted by a vacuum energy in particle physics \cite{Weinberg:1988cp}. Other viable alternatives come from the inclusion of slowly rolling scalar fields which control the accelerated expansion of the Universe at late times, being Quintessence an example of these type of models.

In the framework of string phenomenology, at present we should necessarily be in an epoch of quintessence and in the continuous search for a theory of quantum gravity there has been some Swampland conjectures proposed, which refer to some criteria that must be hold in order to have effective field theories consistent with quantum gravity \cite{Cicoli:2012tz}. Some of these conjectures have important implications in cosmology as is the case of the Swampland criteria that we have analyzed in this paper.

In our analysis we have presented two model independent reconstructions of the Swampland conjectures with ML and cosmography placing constraints on the first and second Swampland conjectures, finding that at low redshifts both approaches give consistent results and that, neither conjecture is ruled out by the data at a statistically significant level.

However, while the swampland bounds seem to be satisfied by the low redshift data, their reconstructed scalar potential, see Fig.~\ref{fig:pot_errkin}, has a shape similar to a parabola. Thus, while these reconstructions may be superficially consistent with the swampland bounds, it should be noted that exponential potentials are what naturally emerges in UV complete settings.

Comparing both of our approaches, the ML algorithm seems a more robust approach for the following reasons: First, in cosmography we are assuming flatness and a Planck 2018 prior on $\Omega_{\mathrm{m},0}$, while with the GA we do not make assumptions on flatness, $\Omega_{\mathrm{m},0}$ or on a dark energy model. Second, our ML reconstructions allow us to place constraints at higher redshifts where the kinetic term reconstructed goes up to $z\sim 1.9$. while for cosmography only extends up to $z\sim 0.3$, as the scalar field kinetic term becomes negative beyond that.

Finally, using the growth rate data $f\sigma_8(z)$ we perform a model independent probe of modified gravity cosmologies through two phase diagrams, $H-f\sigma_8$ and $\eta-f\sigma_8$ where the anisotropic stress parameter $\eta$ is reconstructed through the $E_g$ statistics, related to gravitational lensing data. We see that the first diagram $H-f\sigma_8$ is consistent within the errors with the $\Lambda$CDM model, while the second diagram $\eta-f\sigma_8$ has a $\sim 2\sigma$ deviation of the anisotropic stress from unity at $z\sim 0.3$ and a $\sim 4\sigma$ at $z\sim 0.9$, thus pointing toward mild deviations from GR, which could be the future target of upcoming large-scale structure surveys.

\section*{Acknowledgements}
The authors acknowledge support from the Research Project PGC2018-094773-B-C32 and the Centro de Excelencia Severo Ochoa Program SEV-2016-0597. S.~N. also acknowledges support from the Ram\'{o}n y Cajal program through Grant No. RYC-2014-15843.\\

\textbf{Numerical Analysis Files}: The Genetic Algorithm codes used by the authors in the analysis of the paper can be found at \href{https://github.com/snesseris}{https://github.com/snesseris} and \href{https://github.com/RubenArjona}{https://github.com/RubenArjona}.\\

\begin{appendix} 

\section{Error analysis \label{sec:errors}}

In Eqs.~(\ref{eq:Vx}) and (\ref{eq:phix}) we have rewritten the reconstruction equations for the scalar field in terms of dimensionless variables. Then, by standard error propagation we find that the error for the potential and the kinetic term can be expressed as
\ba
\delta \tilde{V}&=&2E\delta E-\frac{x}{6}\frac{d}{dx}\left(2E\delta E\right)-\frac{1}{2}\delta \Omega_{\mathrm{m},0}x^3,\\
\delta \left(\frac{d \tilde{\phi}}{d x}\right)^{2} &=&\frac{2}{3x}\frac{d}{dx}\Big(\frac{\delta E}{E}\Big)+\frac{2\Omega_{\mathrm{m},0}x\delta E}{E^3}-\frac{x\delta \Omega_{\mathrm{m},0}}{E^2}.~~~~
\ea
Since the Swampland conjecture 2 (SC2) is defined as
\be
\textrm{SC2}=\frac{\left|V_{,\phi}\right|}{V}\equiv \frac{\left|(dV/dz)/(d\phi/dz)\right|}{V}=\frac{\left|V'/\phi'\right|}{V},
\ee
then the error will be given by
\be
\delta \textrm{SC2}=\frac{\left|\left(\frac{\delta V'\phi'-\delta \phi'V'}{\phi'^2}\right)\right| V-\delta V \left|V'/\phi'\right|
}{V^2},
\ee
where the prime is the derivative with respect to the redshift $z$.

\end{appendix}

\bibliographystyle{apsrev4-2}
\bibliography{recon}

\begin{thebibliography}{99}%
\makeatletter
\providecommand \@ifxundefined [1]{%
 \@ifx{#1\undefined}
}%
\providecommand \@ifnum [1]{%
 \ifnum #1\expandafter \@firstoftwo
 \else \expandafter \@secondoftwo
 \fi
}%
\providecommand \@ifx [1]{%
 \ifx #1\expandafter \@firstoftwo
 \else \expandafter \@secondoftwo
 \fi
}%
\providecommand \natexlab [1]{#1}%
\providecommand \enquote  [1]{``#1''}%
\providecommand \bibnamefont  [1]{#1}%
\providecommand \bibfnamefont [1]{#1}%
\providecommand \citenamefont [1]{#1}%
\providecommand \href@noop [0]{\@secondoftwo}%
\providecommand \href [0]{\begingroup \@sanitize@url \@href}%
\providecommand \@href[1]{\@@startlink{#1}\@@href}%
\providecommand \@@href[1]{\endgroup#1\@@endlink}%
\providecommand \@sanitize@url [0]{\catcode `\\12\catcode `\$12\catcode
  `\&12\catcode `\#12\catcode `\^12\catcode `\_12\catcode `\%12\relax}%
\providecommand \@@startlink[1]{}%
\providecommand \@@endlink[0]{}%
\providecommand \url  [0]{\begingroup\@sanitize@url \@url }%
\providecommand \@url [1]{\endgroup\@href {#1}{\urlprefix }}%
\providecommand \urlprefix  [0]{URL }%
\providecommand \Eprint [0]{\href }%
\providecommand \doibase [0]{https://doi.org/}%
\providecommand \selectlanguage [0]{\@gobble}%
\providecommand \bibinfo  [0]{\@secondoftwo}%
\providecommand \bibfield  [0]{\@secondoftwo}%
\providecommand \translation [1]{[#1]}%
\providecommand \BibitemOpen [0]{}%
\providecommand \bibitemStop [0]{}%
\providecommand \bibitemNoStop [0]{.\EOS\space}%
\providecommand \EOS [0]{\spacefactor3000\relax}%
\providecommand \BibitemShut  [1]{\csname bibitem#1\endcsname}%
\let\auto@bib@innerbib\@empty
\bibitem [{\citenamefont {Aghanim}\ \emph {et~al.}(2018)\citenamefont {Aghanim}
  \emph {et~al.}}]{Aghanim:2018eyx}%
  \BibitemOpen
  \bibfield  {author} {\bibinfo {author} {\bibfnamefont {N.}~\bibnamefont
  {Aghanim}} \emph {et~al.} (\bibinfo {collaboration} {Planck}),\ }\href@noop
  {} {\  (\bibinfo {year} {2018})},\ \Eprint {https://arxiv.org/abs/1807.06209}
  {arXiv:1807.06209 [astro-ph.CO]} \BibitemShut {NoStop}%
\bibitem [{\citenamefont {Weinberg}(1989)}]{Weinberg:1988cp}%
  \BibitemOpen
  \bibfield  {author} {\bibinfo {author} {\bibfnamefont {S.}~\bibnamefont
  {Weinberg}},\ }\href {https://doi.org/10.1103/RevModPhys.61.1} {\bibfield
  {journal} {\bibinfo  {journal} {Rev. Mod. Phys.}\ }\textbf {\bibinfo {volume}
  {61}},\ \bibinfo {pages} {1} (\bibinfo {year} {1989})},\ \bibinfo {note}
  {[,569(1988)]}\BibitemShut {NoStop}%
\bibitem [{\citenamefont {Carroll}(2001)}]{Carroll:2000fy}%
  \BibitemOpen
  \bibfield  {author} {\bibinfo {author} {\bibfnamefont {S.~M.}\ \bibnamefont
  {Carroll}},\ }\href {https://doi.org/10.12942/lrr-2001-1} {\bibfield
  {journal} {\bibinfo  {journal} {Living Rev. Rel.}\ }\textbf {\bibinfo
  {volume} {4}},\ \bibinfo {pages} {1} (\bibinfo {year} {2001})},\ \Eprint
  {https://arxiv.org/abs/astro-ph/0004075} {arXiv:astro-ph/0004075 [astro-ph]}
  \BibitemShut {NoStop}%
\bibitem [{\citenamefont {Ratra}\ and\ \citenamefont
  {Peebles}(1988)}]{Ratra:1987rm}%
  \BibitemOpen
  \bibfield  {author} {\bibinfo {author} {\bibfnamefont {B.}~\bibnamefont
  {Ratra}}\ and\ \bibinfo {author} {\bibfnamefont {P.~J.~E.}\ \bibnamefont
  {Peebles}},\ }\href {https://doi.org/10.1103/PhysRevD.37.3406} {\bibfield
  {journal} {\bibinfo  {journal} {Phys. Rev.}\ }\textbf {\bibinfo {volume}
  {D37}},\ \bibinfo {pages} {3406} (\bibinfo {year} {1988})}\BibitemShut
  {NoStop}%
\bibitem [{\citenamefont {Wetterich}(1988)}]{Wetterich:1987fm}%
  \BibitemOpen
  \bibfield  {author} {\bibinfo {author} {\bibfnamefont {C.}~\bibnamefont
  {Wetterich}},\ }\href {https://doi.org/10.1016/0550-3213(88)90193-9}
  {\bibfield  {journal} {\bibinfo  {journal} {Nucl. Phys.}\ }\textbf {\bibinfo
  {volume} {B302}},\ \bibinfo {pages} {668} (\bibinfo {year} {1988})},\ \Eprint
  {https://arxiv.org/abs/1711.03844} {arXiv:1711.03844 [hep-th]} \BibitemShut
  {NoStop}%
\bibitem [{\citenamefont {Caldwell}\ \emph {et~al.}(1998)\citenamefont
  {Caldwell}, \citenamefont {Dave},\ and\ \citenamefont
  {Steinhardt}}]{Caldwell:1997ii}%
  \BibitemOpen
  \bibfield  {author} {\bibinfo {author} {\bibfnamefont {R.~R.}\ \bibnamefont
  {Caldwell}}, \bibinfo {author} {\bibfnamefont {R.}~\bibnamefont {Dave}},\
  and\ \bibinfo {author} {\bibfnamefont {P.~J.}\ \bibnamefont {Steinhardt}},\
  }\href {https://doi.org/10.1103/PhysRevLett.80.1582} {\bibfield  {journal}
  {\bibinfo  {journal} {Phys. Rev. Lett.}\ }\textbf {\bibinfo {volume} {80}},\
  \bibinfo {pages} {1582} (\bibinfo {year} {1998})},\ \Eprint
  {https://arxiv.org/abs/astro-ph/9708069} {arXiv:astro-ph/9708069 [astro-ph]}
  \BibitemShut {NoStop}%
\bibitem [{\citenamefont {Armendariz-Picon}\ \emph {et~al.}(2000)\citenamefont
  {Armendariz-Picon}, \citenamefont {Mukhanov},\ and\ \citenamefont
  {Steinhardt}}]{ArmendarizPicon:2000dh}%
  \BibitemOpen
  \bibfield  {author} {\bibinfo {author} {\bibfnamefont {C.}~\bibnamefont
  {Armendariz-Picon}}, \bibinfo {author} {\bibfnamefont {V.~F.}\ \bibnamefont
  {Mukhanov}},\ and\ \bibinfo {author} {\bibfnamefont {P.~J.}\ \bibnamefont
  {Steinhardt}},\ }\href {https://doi.org/10.1103/PhysRevLett.85.4438}
  {\bibfield  {journal} {\bibinfo  {journal} {Phys. Rev. Lett.}\ }\textbf
  {\bibinfo {volume} {85}},\ \bibinfo {pages} {4438} (\bibinfo {year}
  {2000})},\ \Eprint {https://arxiv.org/abs/astro-ph/0004134}
  {arXiv:astro-ph/0004134 [astro-ph]} \BibitemShut {NoStop}%
\bibitem [{\citenamefont {Armendariz-Picon}\ \emph {et~al.}(2001)\citenamefont
  {Armendariz-Picon}, \citenamefont {Mukhanov},\ and\ \citenamefont
  {Steinhardt}}]{ArmendarizPicon:2000ah}%
  \BibitemOpen
  \bibfield  {author} {\bibinfo {author} {\bibfnamefont {C.}~\bibnamefont
  {Armendariz-Picon}}, \bibinfo {author} {\bibfnamefont {V.~F.}\ \bibnamefont
  {Mukhanov}},\ and\ \bibinfo {author} {\bibfnamefont {P.~J.}\ \bibnamefont
  {Steinhardt}},\ }\href {https://doi.org/10.1103/PhysRevD.63.103510}
  {\bibfield  {journal} {\bibinfo  {journal} {Phys. Rev.}\ }\textbf {\bibinfo
  {volume} {D63}},\ \bibinfo {pages} {103510} (\bibinfo {year} {2001})},\
  \Eprint {https://arxiv.org/abs/astro-ph/0006373} {arXiv:astro-ph/0006373
  [astro-ph]} \BibitemShut {NoStop}%
\bibitem [{\citenamefont {Uzan}(1999)}]{Uzan:1999ch}%
  \BibitemOpen
  \bibfield  {author} {\bibinfo {author} {\bibfnamefont {J.-P.}\ \bibnamefont
  {Uzan}},\ }\href {https://doi.org/10.1103/PhysRevD.59.123510} {\bibfield
  {journal} {\bibinfo  {journal} {Phys. Rev.}\ }\textbf {\bibinfo {volume}
  {D59}},\ \bibinfo {pages} {123510} (\bibinfo {year} {1999})},\ \Eprint
  {https://arxiv.org/abs/gr-qc/9903004} {arXiv:gr-qc/9903004 [gr-qc]}
  \BibitemShut {NoStop}%
\bibitem [{\citenamefont {Perrotta}\ \emph {et~al.}(1999)\citenamefont
  {Perrotta}, \citenamefont {Baccigalupi},\ and\ \citenamefont
  {Matarrese}}]{Perrotta:1999am}%
  \BibitemOpen
  \bibfield  {author} {\bibinfo {author} {\bibfnamefont {F.}~\bibnamefont
  {Perrotta}}, \bibinfo {author} {\bibfnamefont {C.}~\bibnamefont
  {Baccigalupi}},\ and\ \bibinfo {author} {\bibfnamefont {S.}~\bibnamefont
  {Matarrese}},\ }\href {https://doi.org/10.1103/PhysRevD.61.023507} {\bibfield
   {journal} {\bibinfo  {journal} {Phys. Rev.}\ }\textbf {\bibinfo {volume}
  {D61}},\ \bibinfo {pages} {023507} (\bibinfo {year} {1999})},\ \Eprint
  {https://arxiv.org/abs/astro-ph/9906066} {arXiv:astro-ph/9906066 [astro-ph]}
  \BibitemShut {NoStop}%
\bibitem [{\citenamefont {Riazuelo}\ and\ \citenamefont
  {Uzan}(2002)}]{Riazuelo:2001mg}%
  \BibitemOpen
  \bibfield  {author} {\bibinfo {author} {\bibfnamefont {A.}~\bibnamefont
  {Riazuelo}}\ and\ \bibinfo {author} {\bibfnamefont {J.-P.}\ \bibnamefont
  {Uzan}},\ }\href {https://doi.org/10.1103/PhysRevD.66.023525} {\bibfield
  {journal} {\bibinfo  {journal} {Phys. Rev.}\ }\textbf {\bibinfo {volume}
  {D66}},\ \bibinfo {pages} {023525} (\bibinfo {year} {2002})},\ \Eprint
  {https://arxiv.org/abs/astro-ph/0107386} {arXiv:astro-ph/0107386 [astro-ph]}
  \BibitemShut {NoStop}%
\bibitem [{\citenamefont {Dent}\ \emph {et~al.}(2009)\citenamefont {Dent},
  \citenamefont {Stern},\ and\ \citenamefont {Wetterich}}]{Dent:2008vd}%
  \BibitemOpen
  \bibfield  {author} {\bibinfo {author} {\bibfnamefont {T.}~\bibnamefont
  {Dent}}, \bibinfo {author} {\bibfnamefont {S.}~\bibnamefont {Stern}},\ and\
  \bibinfo {author} {\bibfnamefont {C.}~\bibnamefont {Wetterich}},\ }\href
  {https://doi.org/10.1088/1475-7516/2009/01/038} {\bibfield  {journal}
  {\bibinfo  {journal} {JCAP}\ }\textbf {\bibinfo {volume} {0901}},\ \bibinfo
  {pages} {038}},\ \Eprint {https://arxiv.org/abs/0809.4628} {arXiv:0809.4628
  [hep-ph]} \BibitemShut {NoStop}%
\bibitem [{\citenamefont {Amendola}\ and\ \citenamefont
  {Tsujikawa}(2015)}]{Amendola:2015ksp}%
  \BibitemOpen
  \bibfield  {author} {\bibinfo {author} {\bibfnamefont {L.}~\bibnamefont
  {Amendola}}\ and\ \bibinfo {author} {\bibfnamefont {S.}~\bibnamefont
  {Tsujikawa}},\ }\href
  {http://www.cambridge.org/academic/subjects/physics/cosmology-relativity-and-gravitation/dark-energy-theory-and-observations?format=PB&isbn=9781107453982}
  {\emph {\bibinfo {title} {{Dark Energy}}}}\ (\bibinfo  {publisher} {Cambridge
  University Press},\ \bibinfo {year} {2015})\BibitemShut {NoStop}%
\bibitem [{\citenamefont {Tsujikawa}(2013)}]{Tsujikawa:2013fta}%
  \BibitemOpen
  \bibfield  {author} {\bibinfo {author} {\bibfnamefont {S.}~\bibnamefont
  {Tsujikawa}},\ }\href {https://doi.org/10.1088/0264-9381/30/21/214003}
  {\bibfield  {journal} {\bibinfo  {journal} {Class. Quant. Grav.}\ }\textbf
  {\bibinfo {volume} {30}},\ \bibinfo {pages} {214003} (\bibinfo {year}
  {2013})},\ \Eprint {https://arxiv.org/abs/1304.1961} {arXiv:1304.1961
  [gr-qc]} \BibitemShut {NoStop}%
\bibitem [{\citenamefont {'t~Hooft}\ and\ \citenamefont
  {Veltman}(1974)}]{tHooft:1974toh}%
  \BibitemOpen
  \bibfield  {author} {\bibinfo {author} {\bibfnamefont {G.}~\bibnamefont
  {'t~Hooft}}\ and\ \bibinfo {author} {\bibfnamefont {M.}~\bibnamefont
  {Veltman}},\ }\href@noop {} {\bibfield  {journal} {\bibinfo  {journal} {Ann.\
  Inst.\ H.\ Poincare Phys.\ Theor.\ A}\ }\textbf {\bibinfo {volume} {20}},\
  \bibinfo {pages} {69} (\bibinfo {year} {1974})}\BibitemShut {NoStop}%
\bibitem [{\citenamefont {Gubitosi}\ \emph {et~al.}(2013)\citenamefont
  {Gubitosi}, \citenamefont {Piazza},\ and\ \citenamefont
  {Vernizzi}}]{Gubitosi:2012hu}%
  \BibitemOpen
  \bibfield  {author} {\bibinfo {author} {\bibfnamefont {G.}~\bibnamefont
  {Gubitosi}}, \bibinfo {author} {\bibfnamefont {F.}~\bibnamefont {Piazza}},\
  and\ \bibinfo {author} {\bibfnamefont {F.}~\bibnamefont {Vernizzi}},\ }\href
  {https://doi.org/10.1088/1475-7516/2013/02/032} {\bibfield  {journal}
  {\bibinfo  {journal} {JCAP}\ }\textbf {\bibinfo {volume} {02}},\ \bibinfo
  {pages} {032}},\ \Eprint {https://arxiv.org/abs/1210.0201} {arXiv:1210.0201
  [hep-th]} \BibitemShut {NoStop}%
\bibitem [{\citenamefont {Hu}\ \emph {et~al.}(2014)\citenamefont {Hu},
  \citenamefont {Raveri}, \citenamefont {Frusciante},\ and\ \citenamefont
  {Silvestri}}]{Hu:2013twa}%
  \BibitemOpen
  \bibfield  {author} {\bibinfo {author} {\bibfnamefont {B.}~\bibnamefont
  {Hu}}, \bibinfo {author} {\bibfnamefont {M.}~\bibnamefont {Raveri}}, \bibinfo
  {author} {\bibfnamefont {N.}~\bibnamefont {Frusciante}},\ and\ \bibinfo
  {author} {\bibfnamefont {A.}~\bibnamefont {Silvestri}},\ }\href
  {https://doi.org/10.1103/PhysRevD.89.103530} {\bibfield  {journal} {\bibinfo
  {journal} {Phys. Rev. D}\ }\textbf {\bibinfo {volume} {89}},\ \bibinfo
  {pages} {103530} (\bibinfo {year} {2014})},\ \Eprint
  {https://arxiv.org/abs/1312.5742} {arXiv:1312.5742 [astro-ph.CO]}
  \BibitemShut {NoStop}%
\bibitem [{\citenamefont {Arjona}\ \emph
  {et~al.}(2019{\natexlab{a}})\citenamefont {Arjona}, \citenamefont {Cardona},\
  and\ \citenamefont {Nesseris}}]{Arjona:2018jhh}%
  \BibitemOpen
  \bibfield  {author} {\bibinfo {author} {\bibfnamefont {R.}~\bibnamefont
  {Arjona}}, \bibinfo {author} {\bibfnamefont {W.}~\bibnamefont {Cardona}},\
  and\ \bibinfo {author} {\bibfnamefont {S.}~\bibnamefont {Nesseris}},\ }\href
  {https://doi.org/10.1103/PhysRevD.99.043516} {\bibfield  {journal} {\bibinfo
  {journal} {Phys. Rev.}\ }\textbf {\bibinfo {volume} {D99}},\ \bibinfo {pages}
  {043516} (\bibinfo {year} {2019}{\natexlab{a}})},\ \Eprint
  {https://arxiv.org/abs/1811.02469} {arXiv:1811.02469 [astro-ph.CO]}
  \BibitemShut {NoStop}%
\bibitem [{\citenamefont {Arjona}\ \emph
  {et~al.}(2019{\natexlab{b}})\citenamefont {Arjona}, \citenamefont {Cardona},\
  and\ \citenamefont {Nesseris}}]{Arjona:2019rfn}%
  \BibitemOpen
  \bibfield  {author} {\bibinfo {author} {\bibfnamefont {R.}~\bibnamefont
  {Arjona}}, \bibinfo {author} {\bibfnamefont {W.}~\bibnamefont {Cardona}},\
  and\ \bibinfo {author} {\bibfnamefont {S.}~\bibnamefont {Nesseris}},\ }\href
  {https://doi.org/10.1103/PhysRevD.100.063526} {\bibfield  {journal} {\bibinfo
   {journal} {Phys. Rev.}\ }\textbf {\bibinfo {volume} {D100}},\ \bibinfo
  {pages} {063526} (\bibinfo {year} {2019}{\natexlab{b}})},\ \Eprint
  {https://arxiv.org/abs/1904.06294} {arXiv:1904.06294 [astro-ph.CO]}
  \BibitemShut {NoStop}%
\bibitem [{\citenamefont {Arjona}(2020{\natexlab{a}})}]{Arjona:2020gtm}%
  \BibitemOpen
  \bibfield  {author} {\bibinfo {author} {\bibfnamefont {R.}~\bibnamefont
  {Arjona}},\ }in\ \href@noop {} {\emph {\bibinfo {booktitle} {{2nd Hermann
  Minkowski Meeting on the Foundations of Spacetime Physics}}}}\ (\bibinfo
  {year} {2020})\ \Eprint {https://arxiv.org/abs/2010.04764} {arXiv:2010.04764
  [astro-ph.CO]} \BibitemShut {NoStop}%
\bibitem [{\citenamefont {Peebles}\ and\ \citenamefont
  {Ratra}(2003)}]{Peebles:2002gy}%
  \BibitemOpen
  \bibfield  {author} {\bibinfo {author} {\bibfnamefont {P.~J.~E.}\
  \bibnamefont {Peebles}}\ and\ \bibinfo {author} {\bibfnamefont
  {B.}~\bibnamefont {Ratra}},\ }\href
  {https://doi.org/10.1103/RevModPhys.75.559} {\bibfield  {journal} {\bibinfo
  {journal} {Rev. Mod. Phys.}\ }\textbf {\bibinfo {volume} {75}},\ \bibinfo
  {pages} {559} (\bibinfo {year} {2003})},\ \bibinfo {note} {[,592(2002)]},\
  \Eprint {https://arxiv.org/abs/astro-ph/0207347} {arXiv:astro-ph/0207347
  [astro-ph]} \BibitemShut {NoStop}%
\bibitem [{\citenamefont {Heavens}\ \emph {et~al.}(2017)\citenamefont
  {Heavens}, \citenamefont {Fantaye}, \citenamefont {Sellentin}, \citenamefont
  {Eggers}, \citenamefont {Hosenie}, \citenamefont {Kroon},\ and\ \citenamefont
  {Mootoovaloo}}]{Heavens:2017hkr}%
  \BibitemOpen
  \bibfield  {author} {\bibinfo {author} {\bibfnamefont {A.}~\bibnamefont
  {Heavens}}, \bibinfo {author} {\bibfnamefont {Y.}~\bibnamefont {Fantaye}},
  \bibinfo {author} {\bibfnamefont {E.}~\bibnamefont {Sellentin}}, \bibinfo
  {author} {\bibfnamefont {H.}~\bibnamefont {Eggers}}, \bibinfo {author}
  {\bibfnamefont {Z.}~\bibnamefont {Hosenie}}, \bibinfo {author} {\bibfnamefont
  {S.}~\bibnamefont {Kroon}},\ and\ \bibinfo {author} {\bibfnamefont
  {A.}~\bibnamefont {Mootoovaloo}},\ }\href
  {https://doi.org/10.1103/PhysRevLett.119.101301} {\bibfield  {journal}
  {\bibinfo  {journal} {Phys. Rev. Lett.}\ }\textbf {\bibinfo {volume} {119}},\
  \bibinfo {pages} {101301} (\bibinfo {year} {2017})},\ \Eprint
  {https://arxiv.org/abs/1704.03467} {arXiv:1704.03467 [astro-ph.CO]}
  \BibitemShut {NoStop}%
\bibitem [{\citenamefont {Birrell}\ and\ \citenamefont
  {Davies}(1984)}]{Birrell:1982ix}%
  \BibitemOpen
  \bibfield  {author} {\bibinfo {author} {\bibfnamefont {N.}~\bibnamefont
  {Birrell}}\ and\ \bibinfo {author} {\bibfnamefont {P.}~\bibnamefont
  {Davies}},\ }\href {https://doi.org/10.1017/CBO9780511622632} {\emph
  {\bibinfo {title} {{Quantum Fields in Curved Space}}}},\ Cambridge Monographs
  on Mathematical Physics\ (\bibinfo  {publisher} {Cambridge Univ. Press},\
  \bibinfo {address} {Cambridge, UK},\ \bibinfo {year} {1984})\BibitemShut
  {NoStop}%
\bibitem [{\citenamefont {Ooguri}\ and\ \citenamefont
  {Vafa}(2007)}]{Ooguri:2006in}%
  \BibitemOpen
  \bibfield  {author} {\bibinfo {author} {\bibfnamefont {H.}~\bibnamefont
  {Ooguri}}\ and\ \bibinfo {author} {\bibfnamefont {C.}~\bibnamefont {Vafa}},\
  }\href {https://doi.org/10.1016/j.nuclphysb.2006.10.033} {\bibfield
  {journal} {\bibinfo  {journal} {Nucl.\ Phys.\ B}\ }\textbf {\bibinfo {volume}
  {766}},\ \bibinfo {pages} {21} (\bibinfo {year} {2007})},\ \Eprint
  {https://arxiv.org/abs/hep-th/0605264} {arXiv:hep-th/0605264} \BibitemShut
  {NoStop}%
\bibitem [{\citenamefont {Obied}\ \emph {et~al.}(2018)\citenamefont {Obied},
  \citenamefont {Ooguri}, \citenamefont {Spodyneiko},\ and\ \citenamefont
  {Vafa}}]{Obied:2018sgi}%
  \BibitemOpen
  \bibfield  {author} {\bibinfo {author} {\bibfnamefont {G.}~\bibnamefont
  {Obied}}, \bibinfo {author} {\bibfnamefont {H.}~\bibnamefont {Ooguri}},
  \bibinfo {author} {\bibfnamefont {L.}~\bibnamefont {Spodyneiko}},\ and\
  \bibinfo {author} {\bibfnamefont {C.}~\bibnamefont {Vafa}},\ }\href@noop {}
  {\  (\bibinfo {year} {2018})},\ \Eprint {https://arxiv.org/abs/1806.08362}
  {arXiv:1806.08362 [hep-th]} \BibitemShut {NoStop}%
\bibitem [{\citenamefont {Agrawal}\ \emph {et~al.}(2018)\citenamefont
  {Agrawal}, \citenamefont {Obied}, \citenamefont {Steinhardt},\ and\
  \citenamefont {Vafa}}]{Agrawal:2018own}%
  \BibitemOpen
  \bibfield  {author} {\bibinfo {author} {\bibfnamefont {P.}~\bibnamefont
  {Agrawal}}, \bibinfo {author} {\bibfnamefont {G.}~\bibnamefont {Obied}},
  \bibinfo {author} {\bibfnamefont {P.~J.}\ \bibnamefont {Steinhardt}},\ and\
  \bibinfo {author} {\bibfnamefont {C.}~\bibnamefont {Vafa}},\ }\href
  {https://doi.org/10.1016/j.physletb.2018.07.040} {\bibfield  {journal}
  {\bibinfo  {journal} {Phys.\ Lett.\ B}\ }\textbf {\bibinfo {volume} {784}},\
  \bibinfo {pages} {271} (\bibinfo {year} {2018})},\ \Eprint
  {https://arxiv.org/abs/1806.09718} {arXiv:1806.09718 [hep-th]} \BibitemShut
  {NoStop}%
\bibitem [{\citenamefont {Elizalde}\ and\ \citenamefont
  {Khurshudyan}(2019)}]{Elizalde:2018dvw}%
  \BibitemOpen
  \bibfield  {author} {\bibinfo {author} {\bibfnamefont {E.}~\bibnamefont
  {Elizalde}}\ and\ \bibinfo {author} {\bibfnamefont {M.}~\bibnamefont
  {Khurshudyan}},\ }\href {https://doi.org/10.1103/PhysRevD.99.103533}
  {\bibfield  {journal} {\bibinfo  {journal} {Phys.\ Rev.\ D}\ }\textbf
  {\bibinfo {volume} {99}},\ \bibinfo {pages} {103533} (\bibinfo {year}
  {2019})},\ \Eprint {https://arxiv.org/abs/1811.03861} {arXiv:1811.03861
  [astro-ph.CO]} \BibitemShut {NoStop}%
\bibitem [{\citenamefont {Yang}(2020)}]{Yang:2020jze}%
  \BibitemOpen
  \bibfield  {author} {\bibinfo {author} {\bibfnamefont {T.}~\bibnamefont
  {Yang}},\ }\href {https://doi.org/10.1103/PhysRevD.102.083511} {\bibfield
  {journal} {\bibinfo  {journal} {Phys. Rev. D}\ }\textbf {\bibinfo {volume}
  {102}},\ \bibinfo {pages} {083511} (\bibinfo {year} {2020})},\ \Eprint
  {https://arxiv.org/abs/2006.14511} {arXiv:2006.14511 [astro-ph.CO]}
  \BibitemShut {NoStop}%
\bibitem [{\citenamefont {Colg\'ain}\ and\ \citenamefont
  {Yavartanoo}(2019)}]{Colgain:2019joh}%
  \BibitemOpen
  \bibfield  {author} {\bibinfo {author} {\bibfnamefont {E.~O.}\ \bibnamefont
  {Colg\'ain}}\ and\ \bibinfo {author} {\bibfnamefont {H.}~\bibnamefont
  {Yavartanoo}},\ }\href {https://doi.org/10.1016/j.physletb.2019.134907}
  {\bibfield  {journal} {\bibinfo  {journal} {Phys. Lett. B}\ }\textbf
  {\bibinfo {volume} {797}},\ \bibinfo {pages} {134907} (\bibinfo {year}
  {2019})},\ \Eprint {https://arxiv.org/abs/1905.02555} {arXiv:1905.02555
  [astro-ph.CO]} \BibitemShut {NoStop}%
\bibitem [{\citenamefont {Banerjee}\ \emph {et~al.}(2020)\citenamefont
  {Banerjee}, \citenamefont {Cai}, \citenamefont {Heisenberg}, \citenamefont
  {Colg\'ain}, \citenamefont {Sheikh-Jabbari},\ and\ \citenamefont
  {Yang}}]{Banerjee:2020xcn}%
  \BibitemOpen
  \bibfield  {author} {\bibinfo {author} {\bibfnamefont {A.}~\bibnamefont
  {Banerjee}}, \bibinfo {author} {\bibfnamefont {H.}~\bibnamefont {Cai}},
  \bibinfo {author} {\bibfnamefont {L.}~\bibnamefont {Heisenberg}}, \bibinfo
  {author} {\bibfnamefont {E.~O.}\ \bibnamefont {Colg\'ain}}, \bibinfo {author}
  {\bibfnamefont {M.}~\bibnamefont {Sheikh-Jabbari}},\ and\ \bibinfo {author}
  {\bibfnamefont {T.}~\bibnamefont {Yang}},\ }\href@noop {} {\  (\bibinfo
  {year} {2020})},\ \Eprint {https://arxiv.org/abs/2006.00244}
  {arXiv:2006.00244 [astro-ph.CO]} \BibitemShut {NoStop}%
\bibitem [{\citenamefont {Akrami}\ \emph {et~al.}(2019)\citenamefont {Akrami},
  \citenamefont {Kallosh}, \citenamefont {Linde},\ and\ \citenamefont
  {Vardanyan}}]{Akrami:2018ylq}%
  \BibitemOpen
  \bibfield  {author} {\bibinfo {author} {\bibfnamefont {Y.}~\bibnamefont
  {Akrami}}, \bibinfo {author} {\bibfnamefont {R.}~\bibnamefont {Kallosh}},
  \bibinfo {author} {\bibfnamefont {A.}~\bibnamefont {Linde}},\ and\ \bibinfo
  {author} {\bibfnamefont {V.}~\bibnamefont {Vardanyan}},\ }\href
  {https://doi.org/10.1002/prop.201800075} {\bibfield  {journal} {\bibinfo
  {journal} {Fortsch.\ Phys.}\ }\textbf {\bibinfo {volume} {67}},\ \bibinfo
  {pages} {1800075} (\bibinfo {year} {2019})},\ \Eprint
  {https://arxiv.org/abs/1808.09440} {arXiv:1808.09440 [hep-th]} \BibitemShut
  {NoStop}%
\bibitem [{\citenamefont {Raveri}\ \emph {et~al.}(2019)\citenamefont {Raveri},
  \citenamefont {Hu},\ and\ \citenamefont {Sethi}}]{Raveri:2018ddi}%
  \BibitemOpen
  \bibfield  {author} {\bibinfo {author} {\bibfnamefont {M.}~\bibnamefont
  {Raveri}}, \bibinfo {author} {\bibfnamefont {W.}~\bibnamefont {Hu}},\ and\
  \bibinfo {author} {\bibfnamefont {S.}~\bibnamefont {Sethi}},\ }\href
  {https://doi.org/10.1103/PhysRevD.99.083518} {\bibfield  {journal} {\bibinfo
  {journal} {Phys. Rev. D}\ }\textbf {\bibinfo {volume} {99}},\ \bibinfo
  {pages} {083518} (\bibinfo {year} {2019})},\ \Eprint
  {https://arxiv.org/abs/1812.10448} {arXiv:1812.10448 [hep-th]} \BibitemShut
  {NoStop}%
\bibitem [{\citenamefont {Garg}\ and\ \citenamefont
  {Krishnan}(2019)}]{Garg:2018reu}%
  \BibitemOpen
  \bibfield  {author} {\bibinfo {author} {\bibfnamefont {S.~K.}\ \bibnamefont
  {Garg}}\ and\ \bibinfo {author} {\bibfnamefont {C.}~\bibnamefont
  {Krishnan}},\ }\href {https://doi.org/10.1007/JHEP11(2019)075} {\bibfield
  {journal} {\bibinfo  {journal} {JHEP}\ }\textbf {\bibinfo {volume} {11}},\
  \bibinfo {pages} {075}},\ \Eprint {https://arxiv.org/abs/1807.05193}
  {arXiv:1807.05193 [hep-th]} \BibitemShut {NoStop}%
\bibitem [{\citenamefont {Akrami}\ \emph {et~al.}(2020)\citenamefont {Akrami},
  \citenamefont {Sasaki}, \citenamefont {Solomon},\ and\ \citenamefont
  {Vardanyan}}]{Akrami:2020zfz}%
  \BibitemOpen
  \bibfield  {author} {\bibinfo {author} {\bibfnamefont {Y.}~\bibnamefont
  {Akrami}}, \bibinfo {author} {\bibfnamefont {M.}~\bibnamefont {Sasaki}},
  \bibinfo {author} {\bibfnamefont {A.~R.}\ \bibnamefont {Solomon}},\ and\
  \bibinfo {author} {\bibfnamefont {V.}~\bibnamefont {Vardanyan}},\ }\href@noop
  {} {\  (\bibinfo {year} {2020})},\ \Eprint {https://arxiv.org/abs/2008.13660}
  {arXiv:2008.13660 [astro-ph.CO]} \BibitemShut {NoStop}%
\bibitem [{\citenamefont {Heisenberg}\ \emph {et~al.}(2018)\citenamefont
  {Heisenberg}, \citenamefont {Bartelmann}, \citenamefont {Brandenberger},\
  and\ \citenamefont {Refregier}}]{Heisenberg:2018yae}%
  \BibitemOpen
  \bibfield  {author} {\bibinfo {author} {\bibfnamefont {L.}~\bibnamefont
  {Heisenberg}}, \bibinfo {author} {\bibfnamefont {M.}~\bibnamefont
  {Bartelmann}}, \bibinfo {author} {\bibfnamefont {R.}~\bibnamefont
  {Brandenberger}},\ and\ \bibinfo {author} {\bibfnamefont {A.}~\bibnamefont
  {Refregier}},\ }\href {https://doi.org/10.1103/PhysRevD.98.123502} {\bibfield
   {journal} {\bibinfo  {journal} {Phys. Rev. D}\ }\textbf {\bibinfo {volume}
  {98}},\ \bibinfo {pages} {123502} (\bibinfo {year} {2018})},\ \Eprint
  {https://arxiv.org/abs/1808.02877} {arXiv:1808.02877 [astro-ph.CO]}
  \BibitemShut {NoStop}%
\bibitem [{\citenamefont {Sagredo}\ \emph
  {et~al.}(2018{\natexlab{a}})\citenamefont {Sagredo}, \citenamefont
  {Lafaurie},\ and\ \citenamefont {Sapone}}]{Sagredo:2018rvc}%
  \BibitemOpen
  \bibfield  {author} {\bibinfo {author} {\bibfnamefont {B.}~\bibnamefont
  {Sagredo}}, \bibinfo {author} {\bibfnamefont {J.~S.}\ \bibnamefont
  {Lafaurie}},\ and\ \bibinfo {author} {\bibfnamefont {D.}~\bibnamefont
  {Sapone}},\ }\href@noop {} {\  (\bibinfo {year} {2018}{\natexlab{a}})},\
  \Eprint {https://arxiv.org/abs/1808.05660} {arXiv:1808.05660 [astro-ph.CO]}
  \BibitemShut {NoStop}%
\bibitem [{\citenamefont {Moresco}\ and\ \citenamefont
  {Marulli}(2017)}]{Moresco:2017hwt}%
  \BibitemOpen
  \bibfield  {author} {\bibinfo {author} {\bibfnamefont {M.}~\bibnamefont
  {Moresco}}\ and\ \bibinfo {author} {\bibfnamefont {F.}~\bibnamefont
  {Marulli}},\ }\href {https://doi.org/10.1093/mnrasl/slx112} {\bibfield
  {journal} {\bibinfo  {journal} {Mon. Not. Roy. Astron. Soc.}\ }\textbf
  {\bibinfo {volume} {471}},\ \bibinfo {pages} {L82} (\bibinfo {year}
  {2017})},\ \Eprint {https://arxiv.org/abs/1705.07903} {arXiv:1705.07903
  [astro-ph.CO]} \BibitemShut {NoStop}%
\bibitem [{\citenamefont {Basilakos}\ and\ \citenamefont
  {Nesseris}(2017)}]{Basilakos:2017rgc}%
  \BibitemOpen
  \bibfield  {author} {\bibinfo {author} {\bibfnamefont {S.}~\bibnamefont
  {Basilakos}}\ and\ \bibinfo {author} {\bibfnamefont {S.}~\bibnamefont
  {Nesseris}},\ }\href {https://doi.org/10.1103/PhysRevD.96.063517} {\bibfield
  {journal} {\bibinfo  {journal} {Phys. Rev.}\ }\textbf {\bibinfo {volume}
  {D96}},\ \bibinfo {pages} {063517} (\bibinfo {year} {2017})},\ \Eprint
  {https://arxiv.org/abs/1705.08797} {arXiv:1705.08797 [astro-ph.CO]}
  \BibitemShut {NoStop}%
\bibitem [{\citenamefont {Linder}(2017)}]{Linder:2016xer}%
  \BibitemOpen
  \bibfield  {author} {\bibinfo {author} {\bibfnamefont {E.~V.}\ \bibnamefont
  {Linder}},\ }\href {https://doi.org/10.1016/j.astropartphys.2016.11.002}
  {\bibfield  {journal} {\bibinfo  {journal} {Astropart. Phys.}\ }\textbf
  {\bibinfo {volume} {86}},\ \bibinfo {pages} {41} (\bibinfo {year} {2017})},\
  \Eprint {https://arxiv.org/abs/1610.05321} {arXiv:1610.05321 [astro-ph.CO]}
  \BibitemShut {NoStop}%
\bibitem [{\citenamefont {Matsumoto}\ \emph {et~al.}(2020)\citenamefont
  {Matsumoto}, \citenamefont {Okumura},\ and\ \citenamefont
  {Sasaki}}]{Matsumoto:2020kdu}%
  \BibitemOpen
  \bibfield  {author} {\bibinfo {author} {\bibfnamefont {J.}~\bibnamefont
  {Matsumoto}}, \bibinfo {author} {\bibfnamefont {T.}~\bibnamefont {Okumura}},\
  and\ \bibinfo {author} {\bibfnamefont {M.}~\bibnamefont {Sasaki}},\ }\href
  {https://doi.org/10.1088/1475-7516/2020/07/059} {\bibfield  {journal}
  {\bibinfo  {journal} {JCAP}\ }\textbf {\bibinfo {volume} {07}},\ \bibinfo
  {pages} {059}},\ \Eprint {https://arxiv.org/abs/2005.09227} {arXiv:2005.09227
  [astro-ph.CO]} \BibitemShut {NoStop}%
\bibitem [{\citenamefont {Sahni}\ and\ \citenamefont
  {Starobinsky}(2006)}]{Sahni:2006pa}%
  \BibitemOpen
  \bibfield  {author} {\bibinfo {author} {\bibfnamefont {V.}~\bibnamefont
  {Sahni}}\ and\ \bibinfo {author} {\bibfnamefont {A.}~\bibnamefont
  {Starobinsky}},\ }\href {https://doi.org/10.1142/S0218271806009704}
  {\bibfield  {journal} {\bibinfo  {journal} {Int.\ J.\ Mod.\ Phys.\ D}\
  }\textbf {\bibinfo {volume} {15}},\ \bibinfo {pages} {2105} (\bibinfo {year}
  {2006})},\ \Eprint {https://arxiv.org/abs/astro-ph/0610026}
  {arXiv:astro-ph/0610026} \BibitemShut {NoStop}%
\bibitem [{\citenamefont {Scherrer}(2015)}]{Scherrer:2015tra}%
  \BibitemOpen
  \bibfield  {author} {\bibinfo {author} {\bibfnamefont {R.~J.}\ \bibnamefont
  {Scherrer}},\ }\href {https://doi.org/10.1103/PhysRevD.92.043001} {\bibfield
  {journal} {\bibinfo  {journal} {Phys. Rev. D}\ }\textbf {\bibinfo {volume}
  {92}},\ \bibinfo {pages} {043001} (\bibinfo {year} {2015})},\ \Eprint
  {https://arxiv.org/abs/1505.05781} {arXiv:1505.05781 [astro-ph.CO]}
  \BibitemShut {NoStop}%
\bibitem [{\citenamefont {Visser}(2005)}]{Visser:2004bf}%
  \BibitemOpen
  \bibfield  {author} {\bibinfo {author} {\bibfnamefont {M.}~\bibnamefont
  {Visser}},\ }\href {https://doi.org/10.1007/s10714-005-0134-8} {\bibfield
  {journal} {\bibinfo  {journal} {Gen.\ Rel.\ Grav.}\ }\textbf {\bibinfo
  {volume} {37}},\ \bibinfo {pages} {1541} (\bibinfo {year} {2005})},\ \Eprint
  {https://arxiv.org/abs/gr-qc/0411131} {arXiv:gr-qc/0411131} \BibitemShut
  {NoStop}%
\bibitem [{\citenamefont {Capozziello}\ \emph {et~al.}(2008)\citenamefont
  {Capozziello}, \citenamefont {Cardone},\ and\ \citenamefont
  {Salzano}}]{Capozziello:2008qc}%
  \BibitemOpen
  \bibfield  {author} {\bibinfo {author} {\bibfnamefont {S.}~\bibnamefont
  {Capozziello}}, \bibinfo {author} {\bibfnamefont {V.}~\bibnamefont
  {Cardone}},\ and\ \bibinfo {author} {\bibfnamefont {V.}~\bibnamefont
  {Salzano}},\ }\href {https://doi.org/10.1103/PhysRevD.78.063504} {\bibfield
  {journal} {\bibinfo  {journal} {Phys.\ Rev.\ D}\ }\textbf {\bibinfo {volume}
  {78}},\ \bibinfo {pages} {063504} (\bibinfo {year} {2008})},\ \Eprint
  {https://arxiv.org/abs/0802.1583} {arXiv:0802.1583 [astro-ph]} \BibitemShut
  {NoStop}%
\bibitem [{\citenamefont {Capozziello}\ \emph {et~al.}(2019)\citenamefont
  {Capozziello}, \citenamefont {D'Agostino},\ and\ \citenamefont
  {Luongo}}]{Capozziello:2019cav}%
  \BibitemOpen
  \bibfield  {author} {\bibinfo {author} {\bibfnamefont {S.}~\bibnamefont
  {Capozziello}}, \bibinfo {author} {\bibfnamefont {R.}~\bibnamefont
  {D'Agostino}},\ and\ \bibinfo {author} {\bibfnamefont {O.}~\bibnamefont
  {Luongo}},\ }\href {https://doi.org/10.1142/S0218271819300167} {\bibfield
  {journal} {\bibinfo  {journal} {Int.\ J.\ Mod.\ Phys.\ D}\ }\textbf {\bibinfo
  {volume} {28}},\ \bibinfo {pages} {1930016} (\bibinfo {year} {2019})},\
  \Eprint {https://arxiv.org/abs/1904.01427} {arXiv:1904.01427 [gr-qc]}
  \BibitemShut {NoStop}%
\bibitem [{\citenamefont {Aviles}\ \emph {et~al.}(2012)\citenamefont {Aviles},
  \citenamefont {Gruber}, \citenamefont {Luongo},\ and\ \citenamefont
  {Quevedo}}]{Aviles:2012ay}%
  \BibitemOpen
  \bibfield  {author} {\bibinfo {author} {\bibfnamefont {A.}~\bibnamefont
  {Aviles}}, \bibinfo {author} {\bibfnamefont {C.}~\bibnamefont {Gruber}},
  \bibinfo {author} {\bibfnamefont {O.}~\bibnamefont {Luongo}},\ and\ \bibinfo
  {author} {\bibfnamefont {H.}~\bibnamefont {Quevedo}},\ }\href
  {https://doi.org/10.1103/PhysRevD.86.123516} {\bibfield  {journal} {\bibinfo
  {journal} {Phys.\ Rev.\ D}\ }\textbf {\bibinfo {volume} {86}},\ \bibinfo
  {pages} {123516} (\bibinfo {year} {2012})},\ \Eprint
  {https://arxiv.org/abs/1204.2007} {arXiv:1204.2007 [astro-ph.CO]}
  \BibitemShut {NoStop}%
\bibitem [{\citenamefont {Cattoen}\ and\ \citenamefont
  {Visser}(2007)}]{Cattoen:2007id}%
  \BibitemOpen
  \bibfield  {author} {\bibinfo {author} {\bibfnamefont {C.}~\bibnamefont
  {Cattoen}}\ and\ \bibinfo {author} {\bibfnamefont {M.}~\bibnamefont
  {Visser}},\ }\href@noop {} {\  (\bibinfo {year} {2007})},\ \Eprint
  {https://arxiv.org/abs/gr-qc/0703122} {arXiv:gr-qc/0703122} \BibitemShut
  {NoStop}%
\bibitem [{\citenamefont {Lazkoz}\ \emph {et~al.}(2013)\citenamefont {Lazkoz},
  \citenamefont {Alcaniz}, \citenamefont {Escamilla-Rivera}, \citenamefont
  {Salzano},\ and\ \citenamefont {Sendra}}]{Lazkoz:2013ija}%
  \BibitemOpen
  \bibfield  {author} {\bibinfo {author} {\bibfnamefont {R.}~\bibnamefont
  {Lazkoz}}, \bibinfo {author} {\bibfnamefont {J.}~\bibnamefont {Alcaniz}},
  \bibinfo {author} {\bibfnamefont {C.}~\bibnamefont {Escamilla-Rivera}},
  \bibinfo {author} {\bibfnamefont {V.}~\bibnamefont {Salzano}},\ and\ \bibinfo
  {author} {\bibfnamefont {I.}~\bibnamefont {Sendra}},\ }\href
  {https://doi.org/10.1088/1475-7516/2013/12/005} {\bibfield  {journal}
  {\bibinfo  {journal} {JCAP}\ }\textbf {\bibinfo {volume} {12}},\ \bibinfo
  {pages} {005}},\ \Eprint {https://arxiv.org/abs/1311.6817} {arXiv:1311.6817
  [astro-ph.CO]} \BibitemShut {NoStop}%
\bibitem [{\citenamefont {Guimaraes}\ and\ \citenamefont
  {Lima}(2011)}]{Guimaraes:2010mw}%
  \BibitemOpen
  \bibfield  {author} {\bibinfo {author} {\bibfnamefont {A.~C.}\ \bibnamefont
  {Guimaraes}}\ and\ \bibinfo {author} {\bibfnamefont {J.~A.~S.}\ \bibnamefont
  {Lima}},\ }\href {https://doi.org/10.1088/0264-9381/28/12/125026} {\bibfield
  {journal} {\bibinfo  {journal} {Class.\ Quant.\ Grav.}\ }\textbf {\bibinfo
  {volume} {28}},\ \bibinfo {pages} {125026} (\bibinfo {year} {2011})},\
  \Eprint {https://arxiv.org/abs/1005.2986} {arXiv:1005.2986 [astro-ph.CO]}
  \BibitemShut {NoStop}%
\bibitem [{\citenamefont {Gaztanaga}\ \emph {et~al.}(2009)\citenamefont
  {Gaztanaga}, \citenamefont {Cabre},\ and\ \citenamefont
  {Hui}}]{Gaztanaga:2008xz}%
  \BibitemOpen
  \bibfield  {author} {\bibinfo {author} {\bibfnamefont {E.}~\bibnamefont
  {Gaztanaga}}, \bibinfo {author} {\bibfnamefont {A.}~\bibnamefont {Cabre}},\
  and\ \bibinfo {author} {\bibfnamefont {L.}~\bibnamefont {Hui}},\ }\href
  {https://doi.org/10.1111/j.1365-2966.2009.15405.x} {\bibfield  {journal}
  {\bibinfo  {journal} {Mon. Not. Roy. Astron. Soc.}\ }\textbf {\bibinfo
  {volume} {399}},\ \bibinfo {pages} {1663} (\bibinfo {year} {2009})},\ \Eprint
  {https://arxiv.org/abs/0807.3551} {arXiv:0807.3551 [astro-ph]} \BibitemShut
  {NoStop}%
\bibitem [{\citenamefont {Jimenez}\ and\ \citenamefont
  {Loeb}(2002)}]{Jimenez:2001gg}%
  \BibitemOpen
  \bibfield  {author} {\bibinfo {author} {\bibfnamefont {R.}~\bibnamefont
  {Jimenez}}\ and\ \bibinfo {author} {\bibfnamefont {A.}~\bibnamefont {Loeb}},\
  }\href {https://doi.org/10.1086/340549} {\bibfield  {journal} {\bibinfo
  {journal} {Astrophys. J.}\ }\textbf {\bibinfo {volume} {573}},\ \bibinfo
  {pages} {37} (\bibinfo {year} {2002})},\ \Eprint
  {https://arxiv.org/abs/astro-ph/0106145} {arXiv:astro-ph/0106145 [astro-ph]}
  \BibitemShut {NoStop}%
\bibitem [{\citenamefont {Yu}\ \emph {et~al.}(2018)\citenamefont {Yu},
  \citenamefont {Ratra},\ and\ \citenamefont {Wang}}]{Yu:2017iju}%
  \BibitemOpen
  \bibfield  {author} {\bibinfo {author} {\bibfnamefont {H.}~\bibnamefont
  {Yu}}, \bibinfo {author} {\bibfnamefont {B.}~\bibnamefont {Ratra}},\ and\
  \bibinfo {author} {\bibfnamefont {F.-Y.}\ \bibnamefont {Wang}},\ }\href
  {https://doi.org/10.3847/1538-4357/aab0a2} {\bibfield  {journal} {\bibinfo
  {journal} {Astrophys. J.}\ }\textbf {\bibinfo {volume} {856}},\ \bibinfo
  {pages} {3} (\bibinfo {year} {2018})},\ \Eprint
  {https://arxiv.org/abs/1711.03437} {arXiv:1711.03437 [astro-ph.CO]}
  \BibitemShut {NoStop}%
\bibitem [{\citenamefont {Moresco}\ \emph {et~al.}(2016)\citenamefont
  {Moresco}, \citenamefont {Pozzetti}, \citenamefont {Cimatti}, \citenamefont
  {Jimenez}, \citenamefont {Maraston}, \citenamefont {Verde}, \citenamefont
  {Thomas}, \citenamefont {Citro}, \citenamefont {Tojeiro},\ and\ \citenamefont
  {Wilkinson}}]{Moresco:2016mzx}%
  \BibitemOpen
  \bibfield  {author} {\bibinfo {author} {\bibfnamefont {M.}~\bibnamefont
  {Moresco}}, \bibinfo {author} {\bibfnamefont {L.}~\bibnamefont {Pozzetti}},
  \bibinfo {author} {\bibfnamefont {A.}~\bibnamefont {Cimatti}}, \bibinfo
  {author} {\bibfnamefont {R.}~\bibnamefont {Jimenez}}, \bibinfo {author}
  {\bibfnamefont {C.}~\bibnamefont {Maraston}}, \bibinfo {author}
  {\bibfnamefont {L.}~\bibnamefont {Verde}}, \bibinfo {author} {\bibfnamefont
  {D.}~\bibnamefont {Thomas}}, \bibinfo {author} {\bibfnamefont
  {A.}~\bibnamefont {Citro}}, \bibinfo {author} {\bibfnamefont
  {R.}~\bibnamefont {Tojeiro}},\ and\ \bibinfo {author} {\bibfnamefont
  {D.}~\bibnamefont {Wilkinson}},\ }\href
  {https://doi.org/10.1088/1475-7516/2016/05/014} {\bibfield  {journal}
  {\bibinfo  {journal} {JCAP}\ }\textbf {\bibinfo {volume} {1605}}\bibfield
  {number} {\bibinfo  {number} { (05)},\ \bibinfo {pages} {014}},\ }\Eprint
  {https://arxiv.org/abs/1601.01701} {arXiv:1601.01701 [astro-ph.CO]}
  \BibitemShut {NoStop}%
\bibitem [{\citenamefont {Guo}\ and\ \citenamefont
  {Zhang}(2016)}]{Guo:2015gpa}%
  \BibitemOpen
  \bibfield  {author} {\bibinfo {author} {\bibfnamefont {R.-Y.}\ \bibnamefont
  {Guo}}\ and\ \bibinfo {author} {\bibfnamefont {X.}~\bibnamefont {Zhang}},\
  }\href {https://doi.org/10.1140/epjc/s10052-016-4016-x} {\bibfield  {journal}
  {\bibinfo  {journal} {Eur. Phys. J.}\ }\textbf {\bibinfo {volume} {C76}},\
  \bibinfo {pages} {163} (\bibinfo {year} {2016})},\ \Eprint
  {https://arxiv.org/abs/1512.07703} {arXiv:1512.07703 [astro-ph.CO]}
  \BibitemShut {NoStop}%
\bibitem [{\citenamefont {Zhang}\ \emph {et~al.}(2014)\citenamefont {Zhang},
  \citenamefont {Zhang}, \citenamefont {Yuan}, \citenamefont {Zhang},\ and\
  \citenamefont {Sun}}]{Zhang:2012mp}%
  \BibitemOpen
  \bibfield  {author} {\bibinfo {author} {\bibfnamefont {C.}~\bibnamefont
  {Zhang}}, \bibinfo {author} {\bibfnamefont {H.}~\bibnamefont {Zhang}},
  \bibinfo {author} {\bibfnamefont {S.}~\bibnamefont {Yuan}}, \bibinfo {author}
  {\bibfnamefont {T.-J.}\ \bibnamefont {Zhang}},\ and\ \bibinfo {author}
  {\bibfnamefont {Y.-C.}\ \bibnamefont {Sun}},\ }\href
  {https://doi.org/10.1088/1674-4527/14/10/002} {\bibfield  {journal} {\bibinfo
   {journal} {Res. Astron. Astrophys.}\ }\textbf {\bibinfo {volume} {14}},\
  \bibinfo {pages} {1221} (\bibinfo {year} {2014})},\ \Eprint
  {https://arxiv.org/abs/1207.4541} {arXiv:1207.4541 [astro-ph.CO]}
  \BibitemShut {NoStop}%
\bibitem [{\citenamefont {Stern}\ \emph {et~al.}(2010)\citenamefont {Stern},
  \citenamefont {Jimenez}, \citenamefont {Verde}, \citenamefont
  {Kamionkowski},\ and\ \citenamefont {Stanford}}]{STERN:2009EP}%
  \BibitemOpen
  \bibfield  {author} {\bibinfo {author} {\bibfnamefont {D.}~\bibnamefont
  {Stern}}, \bibinfo {author} {\bibfnamefont {R.}~\bibnamefont {Jimenez}},
  \bibinfo {author} {\bibfnamefont {L.}~\bibnamefont {Verde}}, \bibinfo
  {author} {\bibfnamefont {M.}~\bibnamefont {Kamionkowski}},\ and\ \bibinfo
  {author} {\bibfnamefont {S.~A.}\ \bibnamefont {Stanford}},\ }\href
  {https://doi.org/10.1088/1475-7516/2010/02/008} {\bibfield  {journal}
  {\bibinfo  {journal} {JCAP}\ }\textbf {\bibinfo {volume} {1002}},\ \bibinfo
  {pages} {008}},\ \Eprint {https://arxiv.org/abs/0907.3149} {arXiv:0907.3149
  [astro-ph.CO]} \BibitemShut {NoStop}%
\bibitem [{\citenamefont {Moresco}\ \emph {et~al.}(2012)\citenamefont {Moresco}
  \emph {et~al.}}]{MORESCO:2012JH}%
  \BibitemOpen
  \bibfield  {author} {\bibinfo {author} {\bibfnamefont {M.}~\bibnamefont
  {Moresco}} \emph {et~al.},\ }\href
  {https://doi.org/10.1088/1475-7516/2012/08/006} {\bibfield  {journal}
  {\bibinfo  {journal} {JCAP}\ }\textbf {\bibinfo {volume} {1208}},\ \bibinfo
  {pages} {006}},\ \Eprint {https://arxiv.org/abs/1201.3609} {arXiv:1201.3609
  [astro-ph.CO]} \BibitemShut {NoStop}%
\bibitem [{\citenamefont {Chuang}\ and\ \citenamefont
  {Wang}(2013)}]{Chuang:2012qt}%
  \BibitemOpen
  \bibfield  {author} {\bibinfo {author} {\bibfnamefont {C.-H.}\ \bibnamefont
  {Chuang}}\ and\ \bibinfo {author} {\bibfnamefont {Y.}~\bibnamefont {Wang}},\
  }\href {https://doi.org/10.1093/mnras/stt1290} {\bibfield  {journal}
  {\bibinfo  {journal} {Mon. Not. Roy. Astron. Soc.}\ }\textbf {\bibinfo
  {volume} {435}},\ \bibinfo {pages} {255} (\bibinfo {year} {2013})},\ \Eprint
  {https://arxiv.org/abs/1209.0210} {arXiv:1209.0210 [astro-ph.CO]}
  \BibitemShut {NoStop}%
\bibitem [{\citenamefont {Blake}\ \emph {et~al.}(2012)\citenamefont {Blake}
  \emph {et~al.}}]{Blake:2012pj}%
  \BibitemOpen
  \bibfield  {author} {\bibinfo {author} {\bibfnamefont {C.}~\bibnamefont
  {Blake}} \emph {et~al.},\ }\href
  {https://doi.org/10.1111/j.1365-2966.2012.21473.x} {\bibfield  {journal}
  {\bibinfo  {journal} {Mon. Not. Roy. Astron. Soc.}\ }\textbf {\bibinfo
  {volume} {425}},\ \bibinfo {pages} {405} (\bibinfo {year} {2012})},\ \Eprint
  {https://arxiv.org/abs/1204.3674} {arXiv:1204.3674 [astro-ph.CO]}
  \BibitemShut {NoStop}%
\bibitem [{\citenamefont {Anderson}\ \emph {et~al.}(2014)\citenamefont
  {Anderson} \emph {et~al.}}]{Anderson:2013zyy}%
  \BibitemOpen
  \bibfield  {author} {\bibinfo {author} {\bibfnamefont {L.}~\bibnamefont
  {Anderson}} \emph {et~al.} (\bibinfo {collaboration} {BOSS}),\ }\href
  {https://doi.org/10.1093/mnras/stu523} {\bibfield  {journal} {\bibinfo
  {journal} {Mon. Not. Roy. Astron. Soc.}\ }\textbf {\bibinfo {volume} {441}},\
  \bibinfo {pages} {24} (\bibinfo {year} {2014})},\ \Eprint
  {https://arxiv.org/abs/1312.4877} {arXiv:1312.4877 [astro-ph.CO]}
  \BibitemShut {NoStop}%
\bibitem [{\citenamefont {Moresco}(2015)}]{Moresco:2015cya}%
  \BibitemOpen
  \bibfield  {author} {\bibinfo {author} {\bibfnamefont {M.}~\bibnamefont
  {Moresco}},\ }\href {https://doi.org/10.1093/mnrasl/slv037} {\bibfield
  {journal} {\bibinfo  {journal} {Mon. Not. Roy. Astron. Soc.}\ }\textbf
  {\bibinfo {volume} {450}},\ \bibinfo {pages} {L16} (\bibinfo {year}
  {2015})},\ \Eprint {https://arxiv.org/abs/1503.01116} {arXiv:1503.01116
  [astro-ph.CO]} \BibitemShut {NoStop}%
\bibitem [{\citenamefont {Delubac}\ \emph {et~al.}(2015)\citenamefont {Delubac}
  \emph {et~al.}}]{Delubac:2014aqe}%
  \BibitemOpen
  \bibfield  {author} {\bibinfo {author} {\bibfnamefont {T.}~\bibnamefont
  {Delubac}} \emph {et~al.} (\bibinfo {collaboration} {BOSS}),\ }\href
  {https://doi.org/10.1051/0004-6361/201423969} {\bibfield  {journal} {\bibinfo
   {journal} {Astron. Astrophys.}\ }\textbf {\bibinfo {volume} {574}},\
  \bibinfo {pages} {A59} (\bibinfo {year} {2015})},\ \Eprint
  {https://arxiv.org/abs/1404.1801} {arXiv:1404.1801 [astro-ph.CO]}
  \BibitemShut {NoStop}%
\bibitem [{\citenamefont {Sagredo}\ \emph
  {et~al.}(2018{\natexlab{b}})\citenamefont {Sagredo}, \citenamefont
  {Nesseris},\ and\ \citenamefont {Sapone}}]{Sagredo:2018ahx}%
  \BibitemOpen
  \bibfield  {author} {\bibinfo {author} {\bibfnamefont {B.}~\bibnamefont
  {Sagredo}}, \bibinfo {author} {\bibfnamefont {S.}~\bibnamefont {Nesseris}},\
  and\ \bibinfo {author} {\bibfnamefont {D.}~\bibnamefont {Sapone}},\ }\href
  {https://doi.org/10.1103/PhysRevD.98.083543} {\bibfield  {journal} {\bibinfo
  {journal} {Phys. Rev.}\ }\textbf {\bibinfo {volume} {D98}},\ \bibinfo {pages}
  {083543} (\bibinfo {year} {2018}{\natexlab{b}})},\ \Eprint
  {https://arxiv.org/abs/1806.10822} {arXiv:1806.10822 [astro-ph.CO]}
  \BibitemShut {NoStop}%
\bibitem [{\citenamefont {Percival}\ and\ \citenamefont
  {White}(2009)}]{Percival:2008sh}%
  \BibitemOpen
  \bibfield  {author} {\bibinfo {author} {\bibfnamefont {W.~J.}\ \bibnamefont
  {Percival}}\ and\ \bibinfo {author} {\bibfnamefont {M.}~\bibnamefont
  {White}},\ }\href {https://doi.org/10.1111/j.1365-2966.2008.14211.x}
  {\bibfield  {journal} {\bibinfo  {journal} {Mon. Not. Roy. Astron. Soc.}\
  }\textbf {\bibinfo {volume} {393}},\ \bibinfo {pages} {297} (\bibinfo {year}
  {2009})},\ \Eprint {https://arxiv.org/abs/0808.0003} {arXiv:0808.0003
  [astro-ph]} \BibitemShut {NoStop}%
\bibitem [{\citenamefont {Song}\ and\ \citenamefont
  {Percival}(2009)}]{Song:2008qt}%
  \BibitemOpen
  \bibfield  {author} {\bibinfo {author} {\bibfnamefont {Y.-S.}\ \bibnamefont
  {Song}}\ and\ \bibinfo {author} {\bibfnamefont {W.~J.}\ \bibnamefont
  {Percival}},\ }\href {https://doi.org/10.1088/1475-7516/2009/10/004}
  {\bibfield  {journal} {\bibinfo  {journal} {JCAP}\ }\textbf {\bibinfo
  {volume} {0910}},\ \bibinfo {pages} {004}},\ \Eprint
  {https://arxiv.org/abs/0807.0810} {arXiv:0807.0810 [astro-ph]} \BibitemShut
  {NoStop}%
\bibitem [{\citenamefont {Nesseris}\ and\ \citenamefont
  {Perivolaropoulos}(2007)}]{Nesseris:2006er}%
  \BibitemOpen
  \bibfield  {author} {\bibinfo {author} {\bibfnamefont {S.}~\bibnamefont
  {Nesseris}}\ and\ \bibinfo {author} {\bibfnamefont {L.}~\bibnamefont
  {Perivolaropoulos}},\ }\href {https://doi.org/10.1088/1475-7516/2007/01/018}
  {\bibfield  {journal} {\bibinfo  {journal} {JCAP}\ }\textbf {\bibinfo
  {volume} {0701}},\ \bibinfo {pages} {018}},\ \Eprint
  {https://arxiv.org/abs/astro-ph/0610092} {arXiv:astro-ph/0610092 [astro-ph]}
  \BibitemShut {NoStop}%
\bibitem [{\citenamefont {Nesseris}\ and\ \citenamefont
  {Tsujikawa}(2014)}]{Nesseris:2014mea}%
  \BibitemOpen
  \bibfield  {author} {\bibinfo {author} {\bibfnamefont {S.}~\bibnamefont
  {Nesseris}}\ and\ \bibinfo {author} {\bibfnamefont {S.}~\bibnamefont
  {Tsujikawa}},\ }\href {https://doi.org/10.1103/PhysRevD.90.024070} {\bibfield
   {journal} {\bibinfo  {journal} {Phys. Rev.}\ }\textbf {\bibinfo {volume}
  {D90}},\ \bibinfo {pages} {024070} (\bibinfo {year} {2014})},\ \Eprint
  {https://arxiv.org/abs/1402.4613} {arXiv:1402.4613 [astro-ph.CO]}
  \BibitemShut {NoStop}%
\bibitem [{\citenamefont {Nesseris}\ \emph {et~al.}(2017)\citenamefont
  {Nesseris}, \citenamefont {Pantazis},\ and\ \citenamefont
  {Perivolaropoulos}}]{Nesseris:2017vor}%
  \BibitemOpen
  \bibfield  {author} {\bibinfo {author} {\bibfnamefont {S.}~\bibnamefont
  {Nesseris}}, \bibinfo {author} {\bibfnamefont {G.}~\bibnamefont {Pantazis}},\
  and\ \bibinfo {author} {\bibfnamefont {L.}~\bibnamefont {Perivolaropoulos}},\
  }\href {https://doi.org/10.1103/PhysRevD.96.023542} {\bibfield  {journal}
  {\bibinfo  {journal} {Phys. Rev.}\ }\textbf {\bibinfo {volume} {D96}},\
  \bibinfo {pages} {023542} (\bibinfo {year} {2017})},\ \Eprint
  {https://arxiv.org/abs/1703.10538} {arXiv:1703.10538 [astro-ph.CO]}
  \BibitemShut {NoStop}%
\bibitem [{\citenamefont {Kazantzidis}\ and\ \citenamefont
  {Perivolaropoulos}(2018)}]{Kazantzidis:2018rnb}%
  \BibitemOpen
  \bibfield  {author} {\bibinfo {author} {\bibfnamefont {L.}~\bibnamefont
  {Kazantzidis}}\ and\ \bibinfo {author} {\bibfnamefont {L.}~\bibnamefont
  {Perivolaropoulos}},\ }\href {https://doi.org/10.1103/PhysRevD.97.103503}
  {\bibfield  {journal} {\bibinfo  {journal} {Phys. Rev.}\ }\textbf {\bibinfo
  {volume} {D97}},\ \bibinfo {pages} {103503} (\bibinfo {year} {2018})},\
  \Eprint {https://arxiv.org/abs/1803.01337} {arXiv:1803.01337 [astro-ph.CO]}
  \BibitemShut {NoStop}%
\bibitem [{\citenamefont {Arjona}\ \emph
  {et~al.}(2020{\natexlab{a}})\citenamefont {Arjona}, \citenamefont
  {García-Bellido},\ and\ \citenamefont {Nesseris}}]{Arjona:2020yum}%
  \BibitemOpen
  \bibfield  {author} {\bibinfo {author} {\bibfnamefont {R.}~\bibnamefont
  {Arjona}}, \bibinfo {author} {\bibfnamefont {J.}~\bibnamefont
  {García-Bellido}},\ and\ \bibinfo {author} {\bibfnamefont {S.}~\bibnamefont
  {Nesseris}},\ }\href@noop {} {\  (\bibinfo {year} {2020}{\natexlab{a}})},\
  \Eprint {https://arxiv.org/abs/2006.01762} {arXiv:2006.01762 [astro-ph.CO]}
  \BibitemShut {NoStop}%
\bibitem [{\citenamefont {Cardona}\ \emph {et~al.}(2020)\citenamefont
  {Cardona}, \citenamefont {Arjona}, \citenamefont {Estrada},\ and\
  \citenamefont {Nesseris}}]{Cardona:2020ama}%
  \BibitemOpen
  \bibfield  {author} {\bibinfo {author} {\bibfnamefont {W.}~\bibnamefont
  {Cardona}}, \bibinfo {author} {\bibfnamefont {R.}~\bibnamefont {Arjona}},
  \bibinfo {author} {\bibfnamefont {A.}~\bibnamefont {Estrada}},\ and\ \bibinfo
  {author} {\bibfnamefont {S.}~\bibnamefont {Nesseris}},\ }\href@noop {} {\
  (\bibinfo {year} {2020})},\ \Eprint {https://arxiv.org/abs/2012.05282}
  {arXiv:2012.05282 [astro-ph.CO]} \BibitemShut {NoStop}%
\bibitem [{\citenamefont {Arjona}\ and\ \citenamefont
  {Nesseris}(2020)}]{Arjona:2020kco}%
  \BibitemOpen
  \bibfield  {author} {\bibinfo {author} {\bibfnamefont {R.}~\bibnamefont
  {Arjona}}\ and\ \bibinfo {author} {\bibfnamefont {S.}~\bibnamefont
  {Nesseris}},\ }\href@noop {} {\  (\bibinfo {year} {2020})},\ \Eprint
  {https://arxiv.org/abs/2001.11420} {arXiv:2001.11420 [astro-ph.CO]}
  \BibitemShut {NoStop}%
\bibitem [{\citenamefont {Zhang}\ \emph {et~al.}(2007)\citenamefont {Zhang},
  \citenamefont {Liguori}, \citenamefont {Bean},\ and\ \citenamefont
  {Dodelson}}]{Zhang:2007nk}%
  \BibitemOpen
  \bibfield  {author} {\bibinfo {author} {\bibfnamefont {P.}~\bibnamefont
  {Zhang}}, \bibinfo {author} {\bibfnamefont {M.}~\bibnamefont {Liguori}},
  \bibinfo {author} {\bibfnamefont {R.}~\bibnamefont {Bean}},\ and\ \bibinfo
  {author} {\bibfnamefont {S.}~\bibnamefont {Dodelson}},\ }\href
  {https://doi.org/10.1103/PhysRevLett.99.141302} {\bibfield  {journal}
  {\bibinfo  {journal} {Phys. Rev. Lett.}\ }\textbf {\bibinfo {volume} {99}},\
  \bibinfo {pages} {141302} (\bibinfo {year} {2007})},\ \Eprint
  {https://arxiv.org/abs/0704.1932} {arXiv:0704.1932 [astro-ph]} \BibitemShut
  {NoStop}%
\bibitem [{\citenamefont {Reyes}\ \emph {et~al.}(2010)\citenamefont {Reyes},
  \citenamefont {Mandelbaum}, \citenamefont {Seljak}, \citenamefont {Baldauf},
  \citenamefont {Gunn}, \citenamefont {Lombriser},\ and\ \citenamefont
  {Smith}}]{Reyes:2010tr}%
  \BibitemOpen
  \bibfield  {author} {\bibinfo {author} {\bibfnamefont {R.}~\bibnamefont
  {Reyes}}, \bibinfo {author} {\bibfnamefont {R.}~\bibnamefont {Mandelbaum}},
  \bibinfo {author} {\bibfnamefont {U.}~\bibnamefont {Seljak}}, \bibinfo
  {author} {\bibfnamefont {T.}~\bibnamefont {Baldauf}}, \bibinfo {author}
  {\bibfnamefont {J.~E.}\ \bibnamefont {Gunn}}, \bibinfo {author}
  {\bibfnamefont {L.}~\bibnamefont {Lombriser}},\ and\ \bibinfo {author}
  {\bibfnamefont {R.~E.}\ \bibnamefont {Smith}},\ }\href
  {https://doi.org/10.1038/nature08857} {\bibfield  {journal} {\bibinfo
  {journal} {Nature}\ }\textbf {\bibinfo {volume} {464}},\ \bibinfo {pages}
  {256} (\bibinfo {year} {2010})},\ \Eprint {https://arxiv.org/abs/1003.2185}
  {arXiv:1003.2185 [astro-ph.CO]} \BibitemShut {NoStop}%
\bibitem [{\citenamefont {Pinho}\ \emph {et~al.}(2018)\citenamefont {Pinho},
  \citenamefont {Casas},\ and\ \citenamefont {Amendola}}]{Pinho:2018unz}%
  \BibitemOpen
  \bibfield  {author} {\bibinfo {author} {\bibfnamefont {A.~M.}\ \bibnamefont
  {Pinho}}, \bibinfo {author} {\bibfnamefont {S.}~\bibnamefont {Casas}},\ and\
  \bibinfo {author} {\bibfnamefont {L.}~\bibnamefont {Amendola}},\ }\href
  {https://doi.org/10.1088/1475-7516/2018/11/027} {\bibfield  {journal}
  {\bibinfo  {journal} {JCAP}\ }\textbf {\bibinfo {volume} {1811}}\bibfield
  {number} {\bibinfo  {number} { (11)},\ \bibinfo {pages} {027}},\ }\Eprint
  {https://arxiv.org/abs/1805.00027} {arXiv:1805.00027 [astro-ph.CO]}
  \BibitemShut {NoStop}%
\bibitem [{\citenamefont {Skara}\ and\ \citenamefont
  {Perivolaropoulos}(2019)}]{Skara:2019usd}%
  \BibitemOpen
  \bibfield  {author} {\bibinfo {author} {\bibfnamefont {F.}~\bibnamefont
  {Skara}}\ and\ \bibinfo {author} {\bibfnamefont {L.}~\bibnamefont
  {Perivolaropoulos}},\ }\href@noop {} {\  (\bibinfo {year} {2019})},\ \Eprint
  {https://arxiv.org/abs/1911.10609} {arXiv:1911.10609 [astro-ph.CO]}
  \BibitemShut {NoStop}%
\bibitem [{\citenamefont {Bogdanos}\ and\ \citenamefont
  {Nesseris}(2009)}]{Bogdanos:2009ib}%
  \BibitemOpen
  \bibfield  {author} {\bibinfo {author} {\bibfnamefont {C.}~\bibnamefont
  {Bogdanos}}\ and\ \bibinfo {author} {\bibfnamefont {S.}~\bibnamefont
  {Nesseris}},\ }\href {https://doi.org/10.1088/1475-7516/2009/05/006}
  {\bibfield  {journal} {\bibinfo  {journal} {JCAP}\ }\textbf {\bibinfo
  {volume} {0905}},\ \bibinfo {pages} {006}},\ \Eprint
  {https://arxiv.org/abs/0903.2805} {arXiv:0903.2805 [astro-ph.CO]}
  \BibitemShut {NoStop}%
\bibitem [{\citenamefont {Nesseris}\ and\ \citenamefont
  {Shafieloo}(2010)}]{Nesseris:2010ep}%
  \BibitemOpen
  \bibfield  {author} {\bibinfo {author} {\bibfnamefont {S.}~\bibnamefont
  {Nesseris}}\ and\ \bibinfo {author} {\bibfnamefont {A.}~\bibnamefont
  {Shafieloo}},\ }\href {https://doi.org/10.1111/j.1365-2966.2010.17254.x}
  {\bibfield  {journal} {\bibinfo  {journal} {Mon. Not. Roy. Astron. Soc.}\
  }\textbf {\bibinfo {volume} {408}},\ \bibinfo {pages} {1879} (\bibinfo {year}
  {2010})},\ \Eprint {https://arxiv.org/abs/1004.0960} {arXiv:1004.0960
  [astro-ph.CO]} \BibitemShut {NoStop}%
\bibitem [{\citenamefont {Nesseris}\ and\ \citenamefont
  {Garcia-Bellido}(2012)}]{Nesseris:2012tt}%
  \BibitemOpen
  \bibfield  {author} {\bibinfo {author} {\bibfnamefont {S.}~\bibnamefont
  {Nesseris}}\ and\ \bibinfo {author} {\bibfnamefont {J.}~\bibnamefont
  {Garcia-Bellido}},\ }\href {https://doi.org/10.1088/1475-7516/2012/11/033}
  {\bibfield  {journal} {\bibinfo  {journal} {JCAP}\ }\textbf {\bibinfo
  {volume} {1211}},\ \bibinfo {pages} {033}},\ \Eprint
  {https://arxiv.org/abs/1205.0364} {arXiv:1205.0364 [astro-ph.CO]}
  \BibitemShut {NoStop}%
\bibitem [{\citenamefont {Nesseris}\ and\ \citenamefont
  {García-Bellido}(2013)}]{Nesseris:2013bia}%
  \BibitemOpen
  \bibfield  {author} {\bibinfo {author} {\bibfnamefont {S.}~\bibnamefont
  {Nesseris}}\ and\ \bibinfo {author} {\bibfnamefont {J.}~\bibnamefont
  {García-Bellido}},\ }\href {https://doi.org/10.1103/PhysRevD.88.063521}
  {\bibfield  {journal} {\bibinfo  {journal} {Phys. Rev.}\ }\textbf {\bibinfo
  {volume} {D88}},\ \bibinfo {pages} {063521} (\bibinfo {year} {2013})},\
  \Eprint {https://arxiv.org/abs/1306.4885} {arXiv:1306.4885 [astro-ph.CO]}
  \BibitemShut {NoStop}%
\bibitem [{\citenamefont {Sapone}\ \emph {et~al.}(2014)\citenamefont {Sapone},
  \citenamefont {Majerotto},\ and\ \citenamefont {Nesseris}}]{Sapone:2014nna}%
  \BibitemOpen
  \bibfield  {author} {\bibinfo {author} {\bibfnamefont {D.}~\bibnamefont
  {Sapone}}, \bibinfo {author} {\bibfnamefont {E.}~\bibnamefont {Majerotto}},\
  and\ \bibinfo {author} {\bibfnamefont {S.}~\bibnamefont {Nesseris}},\ }\href
  {https://doi.org/10.1103/PhysRevD.90.023012} {\bibfield  {journal} {\bibinfo
  {journal} {Phys. Rev.}\ }\textbf {\bibinfo {volume} {D90}},\ \bibinfo {pages}
  {023012} (\bibinfo {year} {2014})},\ \Eprint
  {https://arxiv.org/abs/1402.2236} {arXiv:1402.2236 [astro-ph.CO]}
  \BibitemShut {NoStop}%
\bibitem [{\citenamefont {Arjona}(2020{\natexlab{b}})}]{Arjona:2020doi}%
  \BibitemOpen
  \bibfield  {author} {\bibinfo {author} {\bibfnamefont {R.}~\bibnamefont
  {Arjona}},\ }\href@noop {} {\  (\bibinfo {year} {2020}{\natexlab{b}})},\
  \Eprint {https://arxiv.org/abs/2002.12700} {arXiv:2002.12700 [astro-ph.CO]}
  \BibitemShut {NoStop}%
\bibitem [{\citenamefont {Arjona}\ and\ \citenamefont
  {Nesseris}(2019)}]{Arjona:2019fwb}%
  \BibitemOpen
  \bibfield  {author} {\bibinfo {author} {\bibfnamefont {R.}~\bibnamefont
  {Arjona}}\ and\ \bibinfo {author} {\bibfnamefont {S.}~\bibnamefont
  {Nesseris}},\ }\href@noop {} {\  (\bibinfo {year} {2019})},\ \Eprint
  {https://arxiv.org/abs/1910.01529} {arXiv:1910.01529 [astro-ph.CO]}
  \BibitemShut {NoStop}%
\bibitem [{\citenamefont {Arjona}\ \emph
  {et~al.}(2020{\natexlab{b}})\citenamefont {Arjona}, \citenamefont {Lin},
  \citenamefont {Nesseris},\ and\ \citenamefont {Tang}}]{Arjona:2020axn}%
  \BibitemOpen
  \bibfield  {author} {\bibinfo {author} {\bibfnamefont {R.}~\bibnamefont
  {Arjona}}, \bibinfo {author} {\bibfnamefont {H.-N.}\ \bibnamefont {Lin}},
  \bibinfo {author} {\bibfnamefont {S.}~\bibnamefont {Nesseris}},\ and\
  \bibinfo {author} {\bibfnamefont {L.}~\bibnamefont {Tang}},\ }\href@noop {}
  {\  (\bibinfo {year} {2020}{\natexlab{b}})},\ \Eprint
  {https://arxiv.org/abs/2011.02718} {arXiv:2011.02718 [astro-ph.CO]}
  \BibitemShut {NoStop}%
\bibitem [{\citenamefont {Nesseris}\ and\ \citenamefont
  {Perivolaropoulos}(2006)}]{Nesseris:2006jc}%
  \BibitemOpen
  \bibfield  {author} {\bibinfo {author} {\bibfnamefont {S.}~\bibnamefont
  {Nesseris}}\ and\ \bibinfo {author} {\bibfnamefont {L.}~\bibnamefont
  {Perivolaropoulos}},\ }\href {https://doi.org/10.1103/PhysRevD.73.103511}
  {\bibfield  {journal} {\bibinfo  {journal} {Phys. Rev. D}\ }\textbf {\bibinfo
  {volume} {73}},\ \bibinfo {pages} {103511} (\bibinfo {year} {2006})},\
  \Eprint {https://arxiv.org/abs/astro-ph/0602053} {arXiv:astro-ph/0602053}
  \BibitemShut {NoStop}%
\bibitem [{\citenamefont {Liddle}(2007)}]{Liddle:2007fy}%
  \BibitemOpen
  \bibfield  {author} {\bibinfo {author} {\bibfnamefont {A.~R.}\ \bibnamefont
  {Liddle}},\ }\href {https://doi.org/10.1111/j.1745-3933.2007.00306.x}
  {\bibfield  {journal} {\bibinfo  {journal} {Mon. Not. Roy. Astron. Soc.}\
  }\textbf {\bibinfo {volume} {377}},\ \bibinfo {pages} {L74} (\bibinfo {year}
  {2007})},\ \Eprint {https://arxiv.org/abs/astro-ph/0701113}
  {arXiv:astro-ph/0701113} \BibitemShut {NoStop}%
\bibitem [{\citenamefont {Nesseris}\ and\ \citenamefont
  {Garcia-Bellido}(2013)}]{Nesseris:2012cq}%
  \BibitemOpen
  \bibfield  {author} {\bibinfo {author} {\bibfnamefont {S.}~\bibnamefont
  {Nesseris}}\ and\ \bibinfo {author} {\bibfnamefont {J.}~\bibnamefont
  {Garcia-Bellido}},\ }\href {https://doi.org/10.1088/1475-7516/2013/08/036}
  {\bibfield  {journal} {\bibinfo  {journal} {JCAP}\ }\textbf {\bibinfo
  {volume} {08}},\ \bibinfo {pages} {036}},\ \Eprint
  {https://arxiv.org/abs/1210.7652} {arXiv:1210.7652 [astro-ph.CO]}
  \BibitemShut {NoStop}%
\bibitem [{\citenamefont {Basilakos}\ \emph {et~al.}(2018)\citenamefont
  {Basilakos}, \citenamefont {Nesseris}, \citenamefont {Anagnostopoulos},\ and\
  \citenamefont {Saridakis}}]{Basilakos:2018arq}%
  \BibitemOpen
  \bibfield  {author} {\bibinfo {author} {\bibfnamefont {S.}~\bibnamefont
  {Basilakos}}, \bibinfo {author} {\bibfnamefont {S.}~\bibnamefont {Nesseris}},
  \bibinfo {author} {\bibfnamefont {F.~K.}\ \bibnamefont {Anagnostopoulos}},\
  and\ \bibinfo {author} {\bibfnamefont {E.~N.}\ \bibnamefont {Saridakis}},\
  }\href {https://doi.org/10.1088/1475-7516/2018/08/008} {\bibfield  {journal}
  {\bibinfo  {journal} {JCAP}\ }\textbf {\bibinfo {volume} {1808}}\bibfield
  {number} {\bibinfo  {number} { (08)},\ \bibinfo {pages} {008}},\ }\Eprint
  {https://arxiv.org/abs/1803.09278} {arXiv:1803.09278 [astro-ph.CO]}
  \BibitemShut {NoStop}%
\bibitem [{\citenamefont {Tr\"oster}\ \emph {et~al.}(2020)\citenamefont
  {Tr\"oster} \emph {et~al.}}]{Troster:2020kai}%
  \BibitemOpen
  \bibfield  {author} {\bibinfo {author} {\bibfnamefont {T.}~\bibnamefont
  {Tr\"oster}} \emph {et~al.},\ }\href@noop {} {\  (\bibinfo {year} {2020})},\
  \Eprint {https://arxiv.org/abs/2010.16416} {arXiv:2010.16416 [astro-ph.CO]}
  \BibitemShut {NoStop}%
\bibitem [{\citenamefont {Multamaki}\ and\ \citenamefont
  {Vilja}(2006)}]{Multamaki:2005zs}%
  \BibitemOpen
  \bibfield  {author} {\bibinfo {author} {\bibfnamefont {T.}~\bibnamefont
  {Multamaki}}\ and\ \bibinfo {author} {\bibfnamefont {I.}~\bibnamefont
  {Vilja}},\ }\href {https://doi.org/10.1103/PhysRevD.73.024018} {\bibfield
  {journal} {\bibinfo  {journal} {Phys. Rev.}\ }\textbf {\bibinfo {volume}
  {D73}},\ \bibinfo {pages} {024018} (\bibinfo {year} {2006})},\ \Eprint
  {https://arxiv.org/abs/astro-ph/0506692} {arXiv:astro-ph/0506692 [astro-ph]}
  \BibitemShut {NoStop}%
\bibitem [{\citenamefont {de~la Cruz-Dombriz}\ and\ \citenamefont
  {Dobado}(2006)}]{delaCruzDombriz:2006fj}%
  \BibitemOpen
  \bibfield  {author} {\bibinfo {author} {\bibfnamefont {A.}~\bibnamefont
  {de~la Cruz-Dombriz}}\ and\ \bibinfo {author} {\bibfnamefont
  {A.}~\bibnamefont {Dobado}},\ }\href
  {https://doi.org/10.1103/PhysRevD.74.087501} {\bibfield  {journal} {\bibinfo
  {journal} {Phys. Rev.}\ }\textbf {\bibinfo {volume} {D74}},\ \bibinfo {pages}
  {087501} (\bibinfo {year} {2006})},\ \Eprint
  {https://arxiv.org/abs/gr-qc/0607118} {arXiv:gr-qc/0607118 [gr-qc]}
  \BibitemShut {NoStop}%
\bibitem [{\citenamefont {Pogosian}\ and\ \citenamefont
  {Silvestri}(2008)}]{Pogosian:2007sw}%
  \BibitemOpen
  \bibfield  {author} {\bibinfo {author} {\bibfnamefont {L.}~\bibnamefont
  {Pogosian}}\ and\ \bibinfo {author} {\bibfnamefont {A.}~\bibnamefont
  {Silvestri}},\ }\href {https://doi.org/10.1103/PhysRevD.77.023503,
  10.1103/PhysRevD.81.049901} {\bibfield  {journal} {\bibinfo  {journal} {Phys.
  Rev.}\ }\textbf {\bibinfo {volume} {D77}},\ \bibinfo {pages} {023503}
  (\bibinfo {year} {2008})},\ \bibinfo {note} {[Erratum: Phys.
  Rev.D81,049901(2010)]},\ \Eprint {https://arxiv.org/abs/0709.0296}
  {arXiv:0709.0296 [astro-ph]} \BibitemShut {NoStop}%
\bibitem [{\citenamefont {Nesseris}(2013)}]{Nesseris:2013fca}%
  \BibitemOpen
  \bibfield  {author} {\bibinfo {author} {\bibfnamefont {S.}~\bibnamefont
  {Nesseris}},\ }\href {https://doi.org/10.1103/PhysRevD.88.123003} {\bibfield
  {journal} {\bibinfo  {journal} {Phys. Rev.}\ }\textbf {\bibinfo {volume}
  {D88}},\ \bibinfo {pages} {123003} (\bibinfo {year} {2013})},\ \Eprint
  {https://arxiv.org/abs/1309.1055} {arXiv:1309.1055 [astro-ph.CO]}
  \BibitemShut {NoStop}%
\bibitem [{\citenamefont {Hu}\ and\ \citenamefont {Sawicki}(2007)}]{Hu:2007nk}%
  \BibitemOpen
  \bibfield  {author} {\bibinfo {author} {\bibfnamefont {W.}~\bibnamefont
  {Hu}}\ and\ \bibinfo {author} {\bibfnamefont {I.}~\bibnamefont {Sawicki}},\
  }\href {https://doi.org/10.1103/PhysRevD.76.064004} {\bibfield  {journal}
  {\bibinfo  {journal} {Phys. Rev. D}\ }\textbf {\bibinfo {volume} {76}},\
  \bibinfo {pages} {064004} (\bibinfo {year} {2007})},\ \Eprint
  {https://arxiv.org/abs/0705.1158} {arXiv:0705.1158 [astro-ph]} \BibitemShut
  {NoStop}%
\bibitem [{\citenamefont {Kimura}\ and\ \citenamefont
  {Yamamoto}(2011)}]{Kimura:2010di}%
  \BibitemOpen
  \bibfield  {author} {\bibinfo {author} {\bibfnamefont {R.}~\bibnamefont
  {Kimura}}\ and\ \bibinfo {author} {\bibfnamefont {K.}~\bibnamefont
  {Yamamoto}},\ }\href {https://doi.org/10.1088/1475-7516/2011/04/025}
  {\bibfield  {journal} {\bibinfo  {journal} {JCAP}\ }\textbf {\bibinfo
  {volume} {04}},\ \bibinfo {pages} {025}},\ \Eprint
  {https://arxiv.org/abs/1011.2006} {arXiv:1011.2006 [astro-ph.CO]}
  \BibitemShut {NoStop}%
\bibitem [{\citenamefont {Arjona}\ \emph
  {et~al.}(2019{\natexlab{c}})\citenamefont {Arjona}, \citenamefont {Cardona},\
  and\ \citenamefont {Nesseris}}]{Arjona:2019zqg}%
  \BibitemOpen
  \bibfield  {author} {\bibinfo {author} {\bibfnamefont {R.}~\bibnamefont
  {Arjona}}, \bibinfo {author} {\bibfnamefont {W.}~\bibnamefont {Cardona}},\
  and\ \bibinfo {author} {\bibfnamefont {S.}~\bibnamefont {Nesseris}},\ }\href
  {https://doi.org/10.1088/1475-7516/2019/10/060} {\bibfield  {journal}
  {\bibinfo  {journal} {JCAP}\ }\textbf {\bibinfo {volume} {10}},\ \bibinfo
  {pages} {060}},\ \Eprint {https://arxiv.org/abs/1906.03160} {arXiv:1906.03160
  [astro-ph.CO]} \BibitemShut {NoStop}%
\bibitem [{\citenamefont {Di~Valentino}\ \emph {et~al.}(2020)\citenamefont
  {Di~Valentino} \emph {et~al.}}]{DiValentino:2020vvd}%
  \BibitemOpen
  \bibfield  {author} {\bibinfo {author} {\bibfnamefont {E.}~\bibnamefont
  {Di~Valentino}} \emph {et~al.},\ }\href@noop {} {\  (\bibinfo {year}
  {2020})},\ \Eprint {https://arxiv.org/abs/2008.11285} {arXiv:2008.11285
  [astro-ph.CO]} \BibitemShut {NoStop}%
\bibitem [{\citenamefont {Muir}\ \emph {et~al.}(2020)\citenamefont {Muir} \emph
  {et~al.}}]{Muir:2020puy}%
  \BibitemOpen
  \bibfield  {author} {\bibinfo {author} {\bibfnamefont {J.}~\bibnamefont
  {Muir}} \emph {et~al.} (\bibinfo {collaboration} {DES}),\ }\href@noop {} {\
  (\bibinfo {year} {2020})},\ \Eprint {https://arxiv.org/abs/2010.05924}
  {arXiv:2010.05924 [astro-ph.CO]} \BibitemShut {NoStop}%
\bibitem [{\citenamefont {Cicoli}\ \emph {et~al.}(2012)\citenamefont {Cicoli},
  \citenamefont {Pedro},\ and\ \citenamefont {Tasinato}}]{Cicoli:2012tz}%
  \BibitemOpen
  \bibfield  {author} {\bibinfo {author} {\bibfnamefont {M.}~\bibnamefont
  {Cicoli}}, \bibinfo {author} {\bibfnamefont {F.~G.}\ \bibnamefont {Pedro}},\
  and\ \bibinfo {author} {\bibfnamefont {G.}~\bibnamefont {Tasinato}},\ }\href
  {https://doi.org/10.1088/1475-7516/2012/07/044} {\bibfield  {journal}
  {\bibinfo  {journal} {JCAP}\ }\textbf {\bibinfo {volume} {07}},\ \bibinfo
  {pages} {044}},\ \Eprint {https://arxiv.org/abs/1203.6655} {arXiv:1203.6655
  [hep-th]} \BibitemShut {NoStop}%
\end{thebibliography}%

\end{document}